\documentclass[10pt,journal,compsoc]{IEEEtran}

\ifCLASSOPTIONcompsoc
  \usepackage[nocompress]{cite}
\else
  \usepackage{cite}
\fi

\usepackage{array}
\usepackage{graphicx}
\usepackage{balance} 
\usepackage{url}
\usepackage{float}
\usepackage{amsmath,graphicx,amsfonts,amssymb,url}
\usepackage{setspace}
\usepackage[caption=false]{subfig}
\usepackage{blindtext}
\usepackage[inline]{enumitem}
\usepackage{color}
 \usepackage{subfig}
\usepackage{multirow}
\newsavebox{\measurebox}
\usepackage{minibox}
\usepackage{array,multirow,booktabs}
\usepackage{graphicx}
\definecolor{mypink1}{rgb}{0.858, 0.188, 0.478}
\usepackage{csquotes}

\ifCLASSINFOpdf
\else
\fi

\begin{document}
%
\title{GANs-NQM: A Generative Adversarial Networks based No Reference Quality Assessment Metric for RGB-D Synthesized Views}
%
%
%
%

\author{Suiyi~Ling,~\IEEEmembership{Student Member,~IEEE,}
Jing~Li, Junle Wang, and~Patrick~Le Callet,~\IEEEmembership{Fellow,~IEEE}
\IEEEcompsocitemizethanks{\IEEEcompsocthanksitem Suiyi Ling, Jing Li, and Patrick Le Callet are with the, group of Image Perception \& Interaction, lab of Laboratory Science Digital De Nantes (LS2N), University of Nantes, Nantes, France.  Email:\{suiyi.ling, patrick.lecallet\}@univ-nantes.fr, jing.li.univ@gmail.com\protect\\
\IEEEcompsocthanksitem Junle Wang is with Tencent, China.
 \IEEEcompsocthanksitem Suiyi Ling and Jing Li make equal contributions. Jing Li is the corresponding author.
}
}


\IEEEtitleabstractindextext{%
\begin{abstract}
In this paper, we proposed a no-reference (NR) quality metric for RGB plus image-depth (RGB-D) synthesis images based on Generative Adversarial Networks (GANs), namely GANs-NQM. Due to the failure of the inpainting on dis-occluded regions in RGB-D synthesis process, to capture the non-uniformly distributed local distortions and to learn their impact on perceptual quality are challenging tasks for objective quality metrics. In our study, based on the characteristics of GANs, we proposed i) a novel training strategy of GANs for RGB-D synthesis images using existing large-scale computer vision datasets rather than RGB-D dataset; ii) a referenceless quality metric based on the trained discriminator by learning a `Bag of Distortion Word' (BDW) codebook and a local distortion regions selector; iii) a hole filling inpainter, i.e., the generator of the trained GANs, for RGB-D dis-occluded regions as a side outcome.  According to the experimental results on IRCCyN/IVC DIBR database, the proposed model outperforms the state-of-the-art quality metrics, in addition, is more applicable in real scenarios.  The corresponding context inpainter also shows appealing results over other inpainting algorithms.
\end{abstract}

}

\maketitle
 
\IEEEdisplaynontitleabstractindextext

%
\IEEEpeerreviewmaketitle

\IEEEraisesectionheading{\section{Introduction}\label{sec:Intro}}
 
\IEEEPARstart{N}{owadays}, thanks to the rapid development of  RGB-D equipment and immersive multi-media technologies, scenarios like 3D-TV~\cite{fehn20033d}, Free-viewpoint TV (FTV)~\cite{tanimoto2011free}, Virtual Reality (VR), and Augmented Reality (AR) have attracted greater users' interests and raised a novel revolution in viewing experience. For instance, FTV offers a `flying in the scene' experience to users by allowing them to change the viewpoints freely in applications like virtual conferences, broadcasting live concerts/matches, remote surveillance,~\textit{etc}. To realize this functionality, new virtual views are needed to be rendered with limited reference videos taken by neighboring reference RGB-D cameras. Since compressing and transmitting massive numbers of RGB-D format images/videos are expensive and inefficient, RGB-D view synthesis is thus important for these use cases to provide users with immersive experience. The depth-image-based rendering (DIBR)~\cite{fehn2004depth} is one of the most widely adopted techniques for synthesizing new views by taking advantages of both the RGB images and the depth maps.

\begin{figure*}[!htbp]
\centering
 \includegraphics[width=1.6\columnwidth]{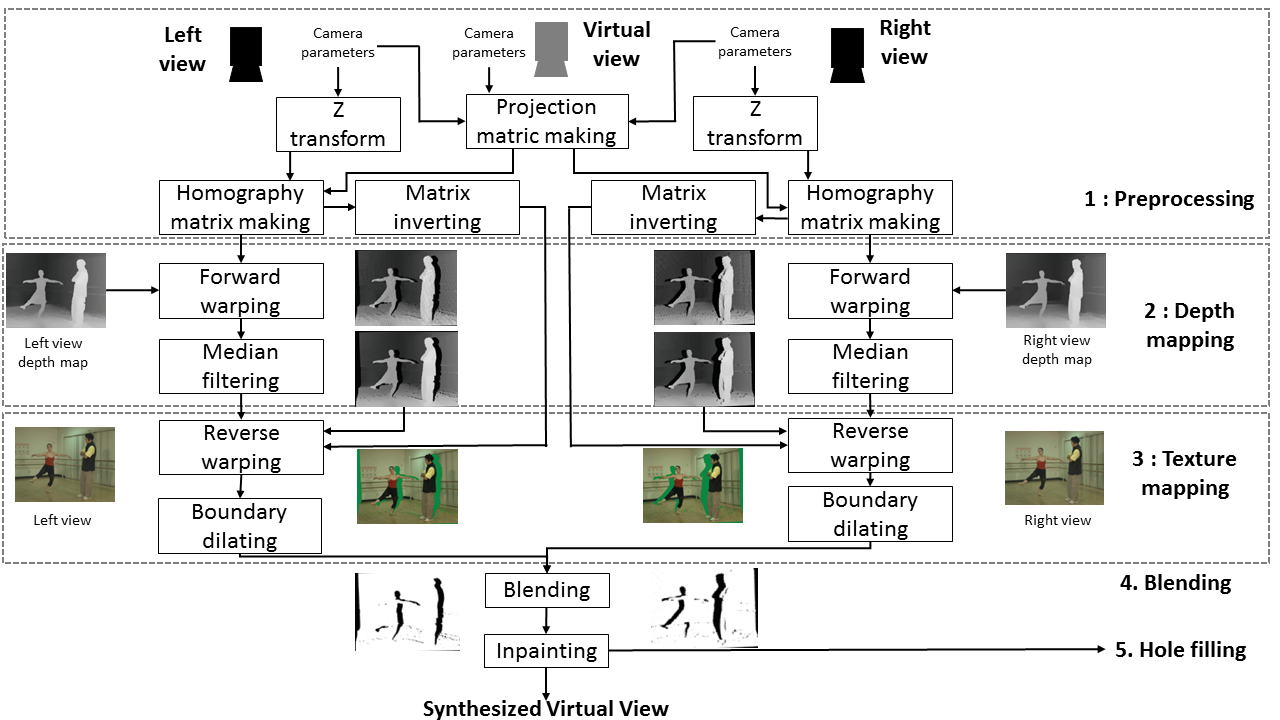}
  \caption{Diagram of DIBR synthesis algorithm. (1) \textbf{Pre-processing procedure}: camera parameters of both the reference and synthesized views are utilized to obtain the projection matrix. (2) \textbf{Depth mapping/virtual depth map generation}, both the left and the right reference depth maps are warped to generate the corresponding virtual depth map by doing forward warping with the transform matrix, \textit{i.e.}, the 3D warping from the reference views to the virtual ones. (3) \textbf{RGB image/texture mapping}, the texture of the virtual view is synthesized by reverse warping, which is to map the texture from the reference views pixel-wise to the virtual ones with the virtual depth maps. (4) \textbf{Blending/`occlusion handling'} process, the left and right synthesized textures are blended together to recover the dis-occluded regions by borrowing information from the two reference views. (5) \textbf{Hole filling process}, inpainting methods are employed to fill up the holes that cannot be handled in the previous steps. }
   \label{fig:framework_DIBR}
\end{figure*}

Nevertheless, the DIBR techniques usually introduce challenging distortions in synthesized views mainly due to occlusions and inaccurate depth maps. A general DIBR based view synthesis scheme could be summarized as a five-step framework as shown in Fig.~\ref{fig:framework_DIBR}, which is proposed by MPEG-FTV~\cite{tanimoto2009view}. Different types of local non-uniform distortions might be introduced in different steps within the DIBR synthesis procedure including local geometric distortion, object shifting/object deformation and most of the time, inpainting related distortions~\cite{bosc2011towards}.

It's worth noting that during the RGB image (texture) mapping process, regions that can be seen in the virtual views but occluded in the reference views are remained as dark holes and commonly termed as dis-occluded regions (`dis-occlusions', or `holes'). Apart from the big dark holes caused by the occlusion, `small holes' may also be introduced by round-off error~\cite{mori2009view} (if pixel coordinates are mapped to an integer value at the virtual viewpoint, then would usually be either interpolated or rounded to their nearest integer position). The `goodness' of an RGB-D synthesized view by using DIBR based techniques highly depends on the hole filling process in these dis-occluded regions, especially when large disocclusions exist~\cite{xu2013depth}.  Therefore, recently, most of the efforts in this domain are spent on developing holes filling algorithms~\cite{yoon2014inter,xu2013depth,ndjiki2011depth,buyssens2015superpixel,buyssens2017depth,gautier2011depth}. However, by employing these inpainting algorithms, `blurry regions' might be induced. Particularly, when it comes to complex texture regions where inpainting algorithms fail to fill up the missing holes, incorrect rendering of texture regions namely `ghosting artifact' might also occur. These blurry or poorly inpainted regions are more visible, since they are always along the foreground background transitions~\cite{battisti2015objective} and sometimes even degrade the structure.

\begin{figure}[!htbp]
\sbox{\measurebox}{%
  \begin{minipage}[b]{0.5\textwidth}
  \subfloat[ ]
 {\label{fig:hole_example_a}\includegraphics[width=0.33\textwidth]{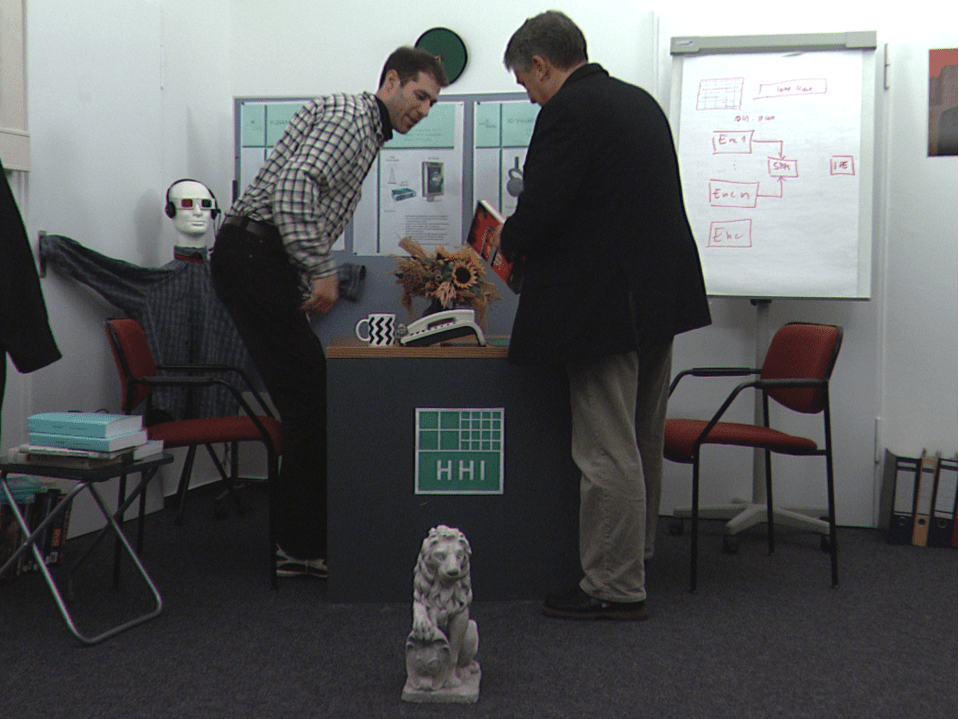}} 
  \subfloat[ ]
 {\label{fig:hole_example_b}\includegraphics[width=0.33\textwidth]{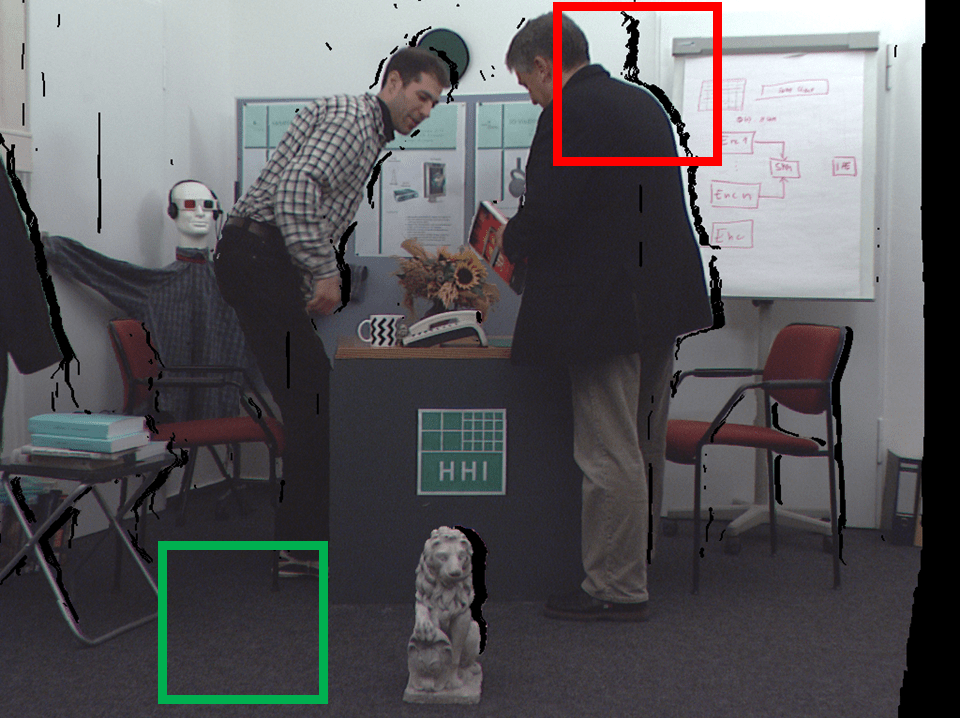}} 
     \subfloat[ ]
  {\label{fig:hole_example_c}\includegraphics[width=0.33\textwidth]{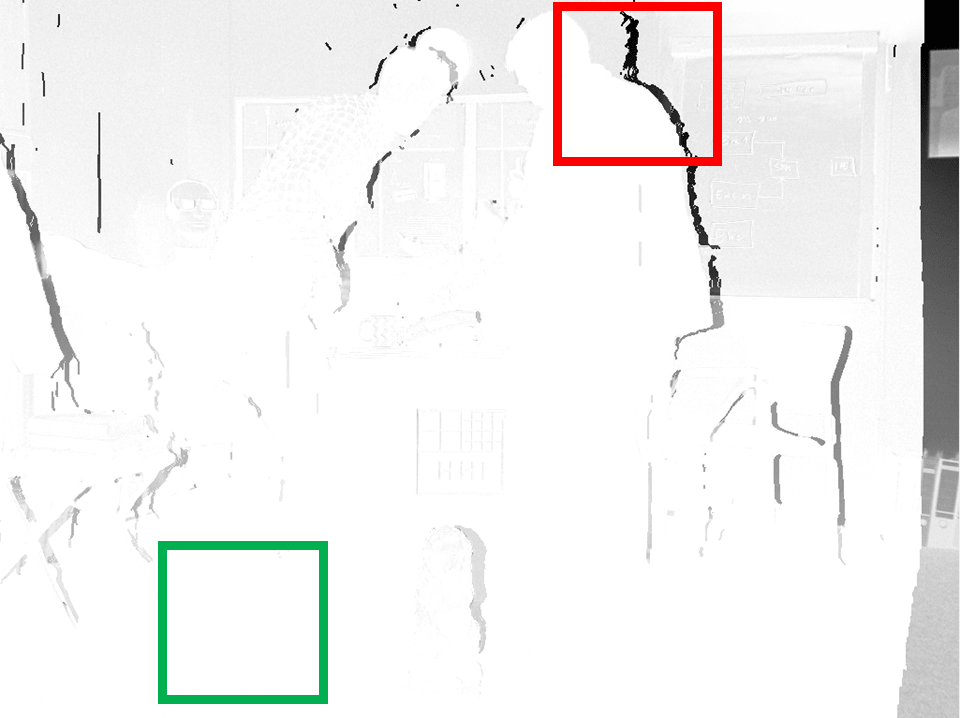}}
  \\
 \subfloat[  ]
  {\label{fig:hole_example_d}\includegraphics[width=0.2\textwidth]{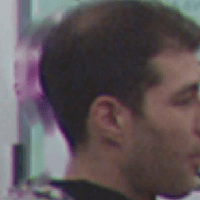}}
    \subfloat[ ]
  {\label{fig:hole_example_e}\includegraphics[width=0.2\textwidth]{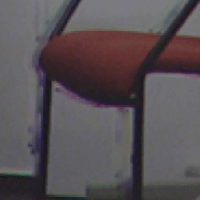}}
    \subfloat[ ]
  {\label{fig:hole_example_f}\includegraphics[width=0.2\textwidth]{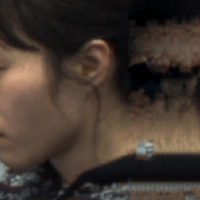}}
    \subfloat[ ]
  {\label{fig:hole_example_g}\includegraphics[width=0.2\textwidth]{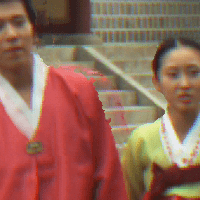}} 
      \subfloat[ ]
  {\label{fig:hole_example_h}\includegraphics[width=0.2\textwidth]{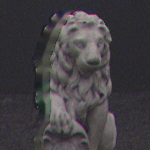}} 
\end{minipage}}
\usebox{\measurebox}\qquad
\caption{ Examples of dis-occluded regions and non-uniform inpainted distortions in DIBR based synthesized views. (a) the reference view. (b) synthesized view with dis-occluded regions. (c) error map between reference and synthesized image (the darker the color, the more distortions there are). Green box region is with non-perceivable distortion while red box region is with big black holes. (d)-(h) examples of local non-uniform inpainted distortions.}
\label{Fig:example_nonuniform}
\end{figure}

Examples of dis-occluded regions and non-uniform inpainted distortions are shown in Fig.\ref{Fig:example_nonuniform}.  Fig. \ref{fig:hole_example_a} is a reference image, the corresponding synthesized view is shown in Fig.~\ref{fig:hole_example_b}, where the dis-occluded regions are distributed locally and non-uniformly as shown in the error map Fig.~\ref{fig:hole_example_c}. After inpainting the dis-occluded regions, local disturbing inpainting artifacts could be observed, examples are shown in Fig.~\ref{fig:hole_example_d} - \ref{fig:hole_example_h}.

Quality control of the entire multi-view display system is important in order to provide users with qualified service. For instance, in the FTV system, synthesis process could be the `bottleneck' of delivering good quality of free viewing experience in extreme cases. Unlike traditional uniform compression distortions, local non-uniform structure related distortions introduced by DIBR synthesis algorithms could be predominant in impacting the perceived quality. The entire viewing experience of a Free Viewpoint Video (FVV) could be ruined by only one poor-quality region in one synthesized view~\cite{gu2017model} in extreme cases. Therefore, a robust quality assessment metric is necessary to benchmark DIBR based synthesized algorithms, provide guidance for the optimization of the overall system performance and improve the Quality of Experience (QoE) of the users. However, these DIBR related artifacts, especially the ones introduced by dis-occluded regions filling, are challenging for existing commonly used quality metrics~\cite{gu2017model,bosc2011towards} as they are designed for the uniform distributed artifacts. A specific model, which can be used to capture the local non-uniformed inpainting distortions in RGB-D view synthesis process is needed. Targeting at solving this issue, we seek a state-of-the-art machine learning technique by asking ourselves \textbf{`How to detect the non-uniform inpainting artifacts without reference and eventually to learn the impact of them on the perceived quality?'} 


Generative Adversarial Network (GANs), first proposed by Goodfellow~\textit{et al.} \cite{goodfellow2014generative}, might be an answer to this question. GANs has been widely used in solving problems in different domains, including 3D structure reconstruction~\cite{yang2018dense}, semantic inpainting (~\textit{i.e.}, context encoder) \cite{wang2017shape}\cite{yeh2017semantic}\cite{pathak2016context}, face recognition~\cite{zhao20183d}, realistic image synthesis~\cite{zhang2017stackgan++}, \textit{etc}. The main idea of the adversarial nets framework is to train a generator ($G$) and a discriminator ($D$) simultaneously: a generative model that captures the data distribution and a discriminative model  \cite{goodfellow2014generative} that is able to tell the `real' image from the generated one. They are trained together so that the discriminator can keep pitting against the generator until the counterfeits generated by the generator cannot be distinguished by the discriminator. By doing so, both of them would be driven to improve their performance until the probability that $D$ makes a mistake is maximized. Taking context encoder as an example, the goal of $G$ is to inpaint an image, with a mask map indicating the regions to be inpainted, to be `real' for the discriminator; the goal of $D$ is being able to distinguish between an inpainted image and a real image,~\textit{i.e.}, ground truth image.  

Considering the case of synthesizing RGB-D views by DIBR techniques, the most annoying non-uniform distortions usually are the non-continuous inpainted distortions introduced by the hole filling stage. If the context encoder is trained to inpaint dis-occluded regions as shown in Fig.~\ref{fig:hole_example_b}, then the discriminator, which is trained along with the generator, could tell whether or not the input image is inpainted. Furthermore, if dividing the input image into local patches, the local inpainted regions could be detected by the trained discriminator, and the perceptual quality of the whole image could also be learned in some way.
 
Based on this assumption, in this paper, a NR quality metric is proposed for evaluating RGB-D synthesized images with the following contributions:  1) A novel RGB-D synthesis view training strategy for deep neural networks (DNNs) is proposed, which is a GANs based context inpainter by designing special masks similar to~\textbf{dis-occluded regions} induced in the general DIBR process. As a by-product, the trained context inpainter (generator) could also be used in DIBR framework for dis-occluded regions inpainting; 2) A local non-uniform distortion region detection strategy is proposed based on the pre-trained discriminator; 3) A quality-aware `Bag of Distortion Words' is learned from high-level features of discriminator to obtain a novel quality-relevant representation for each synthesized image. The source code is publicly available in Github.

The remainder of this paper is organized as follows. The related works including the state-of-the-art quality metrics designed for RGB-D synthesized images and details of GANs based context inpainter are introduced in section~\ref{sec:RW}. In Section~\ref{sec:PM}, the proposed model is introduced in detail. Then, the experimental results and analysis are presented in section \ref{sec:exp}. Finally, conclusions are given in Section~\ref{sec:con}.
  
\section{Related Work}
\label{sec:RW}

\subsection{Conventional image quality metrics not applicable to RGB-D synthesized images}
There are numerous image quality metrics~\cite{chandler2013seven} designed for evaluating the uniformly distributed distortions, for instance, blurriness and blockiness induced by different compression technologies. These metrics have achieved a big success which show high consistency with human perception. However, most of these conventional quality metrics including PSNR, SSIM \cite{wang2004image} fail to well predict the perceived quality of RGB-D synthesized views for the following reasons:
\begin{enumerate}
\item Point-wise metrics like PSNR over-penalize the acceptable global uniform `object shifting' artifact due to the mis-matched correspondences. In another word, the global shifting of objects that may not be noticed by human observers could be penalized heavily by this type of metrics.

\item Most of the existing quality metrics are not designed for local non-uniform artifacts, such as the inpainted distortions. While observing an image, the artifacts located at a `regions of interest' are much more annoying than the one located at inconspicuous areas~\cite{ninassi2007does}. Meanwhile, regions of `poor quality' are more likely to be perceived by humans in an image with more severity than the ones with `good quality'. Thus, images with even a small number of `poor quality' regions are penalized more gravely. In the case of DIBR synthesis views, most of the geometric and inpainting distortion regions locate along the boundaries of `Regions of Interest'~\cite{le2010overt}, which makes them easier to be noticed by observers and leads to bad quality judgments.  

\item In practice, considering the fact that the reference of the virtual RGB-D views are generally not available, a well-designed no-reference (NR) quality metric, which is capable of quantifying those local synthesized related distortions on perceptual quality, is in urgent demand. In the wake of development in machine learning technologies, many NR quality assessment models have been proposed recently on the base of advanced deep learning schemes. However, most of them are proposed based on the assumption that the perceived quality of local regions is the same with the perceptual quality of the entire image~\cite{CNN_metric,bosse2016deep,li2017exploiting}. This assumption may (in most cases, may not) work for images containing uniform distortions but may not stand for those which contain non-uniform distortions as shown in Fig. \ref{Fig:example_nonuniform} since severely distorted local regions are more disturbing and greatly affect the perceived quality of the entire image~\cite{ninassi2007does}.
\end{enumerate}

\subsection{Existing quality metrics for RGB-D synthesized images}
\label{sec:3DM}
To resolve the issues mentioned above, objective quality assessment metrics designed for RGB-D synthesized views are developed, which can be generally classified into full reference (FR) metric and NR metric. Among the FR metrics, one of the very first ones is VSQA \cite{conze2012objective}, which improves SSIM with three visibility maps by characterizing the complexity of the images. The 3DswIM is proposed by Battisti~\textit{et al.} \cite{battisti2015objective} based on statistical features of wavelet sub-bands. Stankovi{\'c}  \cite{sandic2015dibr} first employs morphological wavelet decomposition for quality assessment of synthesized images, namely MW-PSNR. Later, another metric, which devises PSNR with morphological pyramids decomposition (MP-PSNR), is proposed in \cite{sandic2015dibrMP}. Targeting the problem that global shifting artifact is normally over-penalized by point-wise metrics, CT-IQM \cite{CT_IQA} is proposed using a context tree based encoding scheme. To quantify the change of contours' categories from a higher level, ST-IQM is proposed in~\cite{ling2017image} using Sketch Token descriptor.  To quantify the deformation of curves in synthesized views, EM-IQM is proposed in~\cite{EM-IQA} based on an elastic metric.  Li~\textit{et al.}~\cite{li2018quality} proposed LOGs by considering both the geometric distortions as well as the sharpness of the images. Nevertheless, since the references of the synthesized views are generally not available, NR metric is more desirable. Compared to FR metrics mentioned above, only a few NR metrics are designed for synthesized views. In \cite{tian2017niqsv}, NIQSV is proposed based on a strong hypothesis that high-quality images are consist of flat areas separated by edges. Later on, NIQSV+ is introduced in \cite{tian2018niqsv+} to improve NIQSV by taking `black holes' into account. Recently, a novel NR quality metric of DIBR-synthesized images is proposed in \cite{gu2017model} using the auto-regression (AR) based local image description. However, all of these metrics still suffer at least one of the following issues: 1) over penalizing uniform shifting; 2) underestimating local non-uniform inpainted distortions; 3) high computational complexity.

\section{The Proposed Model}
\label{sec:PM}
In this section, the proposed GANs based No-reference Quality Metric for synthesized views (GANs-NQM) is described in detail.

\begin{figure*}[!htbp]
\centering
 \includegraphics[width=1.8\columnwidth]{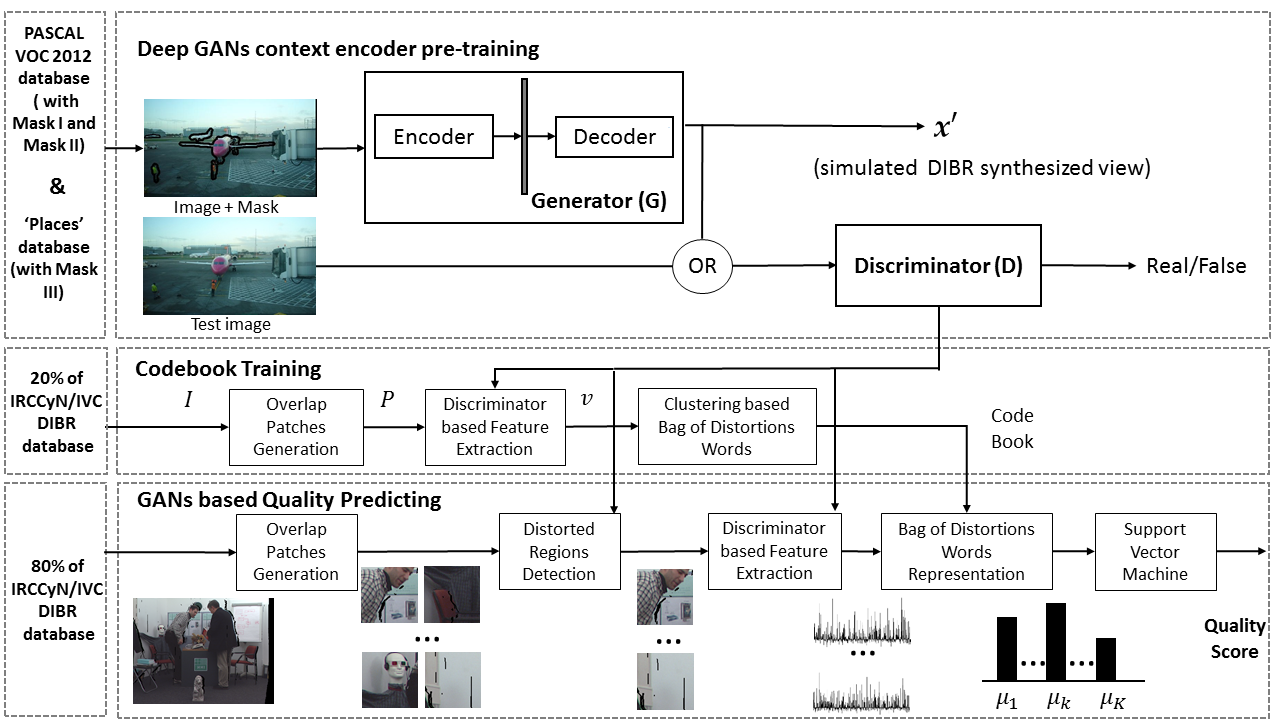}
  \caption{ Diagram of the proposed model: (1) Deep GANs context encoder pre-training; (2) Distortion codebook training; (3) Quality predicting. } 
   \label{fig:framework}
\end{figure*}
 
First and foremost, to train a deep neural network (in our case, GANs), it needs a large number of data samples. Even though there are already many existing RGB-D databases~\cite{firman2016rgbd} publicly available for researchers targeting at computer vision tasks, the diversity of contents are quite limited compared to the general 2D image datasets such as ImageNet~\cite{deng2009imagenet}, CIFAR10/100~\cite{krizhevsky2009learning}, PASCAL VOC~\cite{everingham2015pascal}, and Places challenge~\cite{zhou2017places}. Further considering the fact that the corresponding quality of depth data in RGB-D datasets are generally noisy, in this paper, we propose a novel training strategy for RGB-D synthesis view without using RGB-D database (RGB image and depth map) but designing special masks to mimic the holes that locate at possible dis-occluded regions or possible distorted regions introduced in DIBR process. 

In our proposed model, a generator is trained to impaint images with holes that are similar to the synthesized views generated by DIBR based methodologies, the discriminator is thus to capture the quality information of the RGB-D synthesized views. The entire scheme is depicted in Fig. \ref{fig:framework} correspondingly. Details of each procedure are given in the following subsections.

\subsection{ Simulation the process of inpainting dis-occluded regions using GANs}

\subsubsection{Semantic image inpainting based on GANs}

In the field of computer vision, semantic inpainting is a brand new application, where the goal is to infer the missing regions within an image according to its semantics. Unlike traditional inpainting or texture synthesis methodologies, semantic inpainting \cite{liu2016learning,ren2015shepard,xie2012image,mairal2008sparse, yeh2017semantic, pathak2016context} aims at filling the missing parts by using statistical information from external dataset instead of only making use of the internal property of the image needed to be inpainted.

Among the existing state-of-the-art semantic context inpainters, the ones proposed in \cite{yeh2017semantic,pathak2016context} based on Generative Adversarial Networks (GANs) provide the most appealing performance. In both works, the proposed context encoder (generator) is designed as an auto-encoder with an unfilled image as input. In detail, to insure continuity within the context, $L_2$ norm reconstruction is defined in Equation~(\ref{eq:l_rec}) to regress the missing parts to the ground truth content: 
\begin{equation}
\label{eq:l_rec}
\mathcal{L}_{rec}(x)= \parallel M \odot (x - G( (1-M) \odot x ) ) \parallel _2^2
\end{equation}
where $M$ denotes the binary mask indicating the missing regions need to be inpainted. To overcome the blurry preference problem aroused by $L_2$ loss, \textit{i.e.}, it tends to predict the mean of the distribution resulting in an averaged blurrier image, the adversarial loss is introduced to jointly optimize both $G$ and $D$ as formalized in equation (\ref{eq:l_joint}): 
\begin{equation}
\label{eq:l_joint}
\mathcal{L}_{joint} = \lambda\mathcal{L}_{rec} + (1-\lambda)\mathcal{L}_{adv}
\end{equation}
where $\lambda$ is a hyper-parameter balancing the weights between the two losses and $\mathcal{L}_{adv}$ is further defined in (\ref{l_adv}) by customizing GANs for the context encoder task with the mask $M$:

\begin{equation}
\label{l_adv}
 \mathcal{L}_{adv}= \underset{D}{max} \ \mathbb{E}_{x\sim p_{\textbf{x}}}[log(D(x)) + log(1-D(G( (1-M) \odot x )))]
\end{equation}

In this paper, based on the similar recipe, we design our own masks $M$, which mimics the `dis-occluded' regions that appear during DIBR process, to retrain a new `context inpainter'. Then, we explore to use the pre-train discriminator to evaluate the quality of RGB-D synthesized images.

\subsubsection{Design of masks}
 
As introduced before, dis-occluded regions are introduced during the DIBR synthesis process. There are mainly two types of dis-occluded regions: 1) edge-like holes that are located along the boundaries of the foreground objects as shown in Fig. \ref{fig:example_holes1}, and 2) small or medium size of holes that are distributed throughout the entire images as shown in Fig. \ref{fig:example_holes2}. The shapes of these regions are normally related to the shapes of objects. These regions can be filled with certain inpainting algorithms. However, inpainting-related artifacts may also be introduced.

Generally, dis-occluded regions that are located along the border between the foreground and the background are challenging for existing inpainting algorithms. It is often to see that foreground regions are inpainted with background pixels or vice versa. As a result, the structures of objects are disrupted. Structure related degradation around foreground objects, accompanying with inter-view inconsistency on depth, might then cause binocular rivalry, binocular suppression, or binocular superposition~\cite{jung2011quantitative,
li2013methods} which eventually lead to visual discomfort. Concerning the issues above, and to train a new context encoder that is capable of inpainting the distortions mentioned above, two types of masks $M$ are designed:

\begin{figure}[!htbp]
\sbox{\measurebox}{%
  \begin{minipage}[b]{0.24\textwidth}
   \subfloat[]
  {\label{fig:example_holes1}\includegraphics[width=0.99\textwidth]{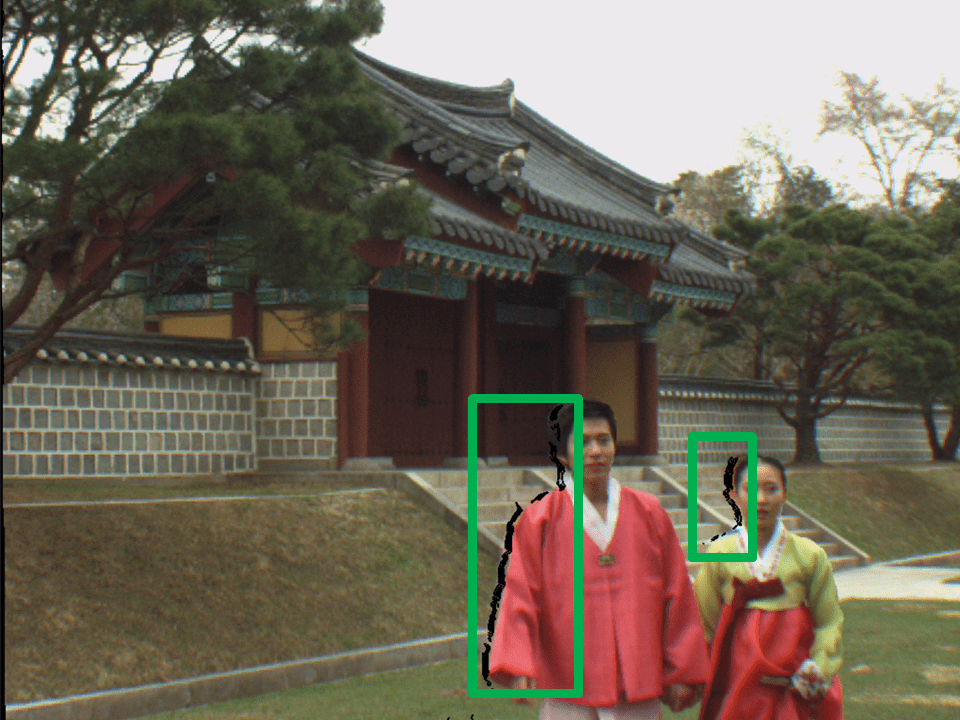}} 
\end{minipage} 
  \begin{minipage}[b]{0.24\textwidth}
  \subfloat[]
  {\label{fig:example_holes2}\includegraphics[width=\textwidth]{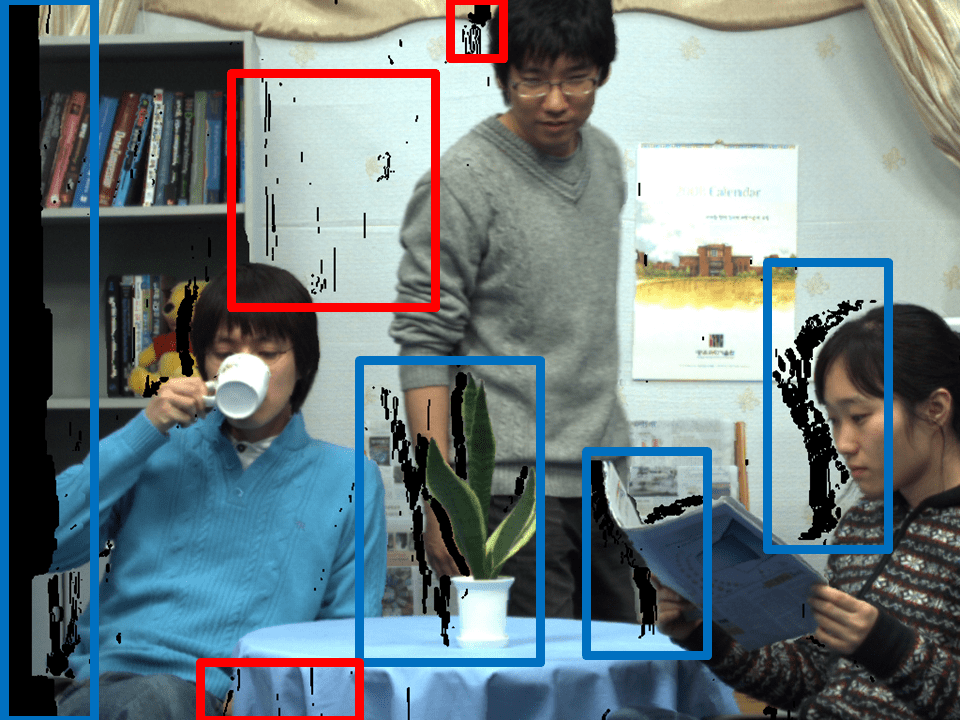}}
\end{minipage}}
\usebox{\measurebox}\qquad
\caption{Examples of typical dis-occluded regions introduced during the process of DIBR based views synthesis. (a) Examples of dis-occluded regions that are around foreground objects' boundaries (bounded by green boxes); (b) Examples of small and median size of dis-occluded regions  (bounded by red and blue bounding boxes correspondingly) that distributed throughout the image;}
\label{fig:example_DOR}
\end{figure}

\begin{figure}[!htbp]
\centering
\subfloat[ Image ]{\label{fig:original_img}
\includegraphics[width=0.12\textwidth]{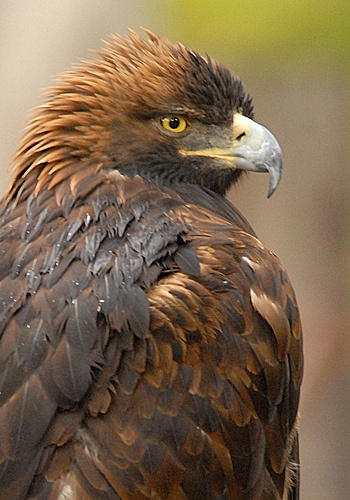}}
\subfloat[Labeled map ]{\label{fig:labeled_map}
\includegraphics[width=0.12\textwidth]{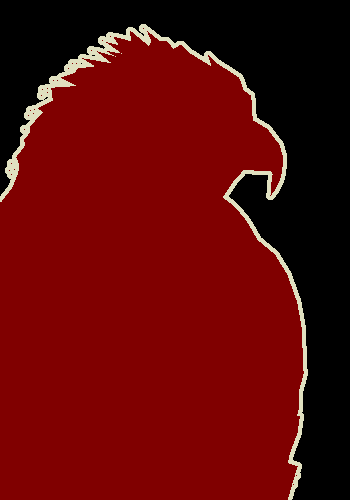}}
\subfloat[Mask I]{\label{fig:maskI}
\includegraphics[width=0.12\textwidth]{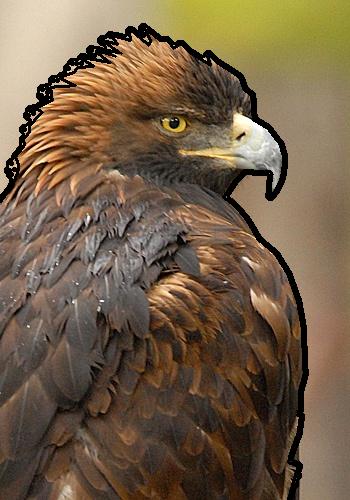}}
\subfloat[Mask II]{\label{fig:maskII}
\includegraphics[width=0.12\textwidth]{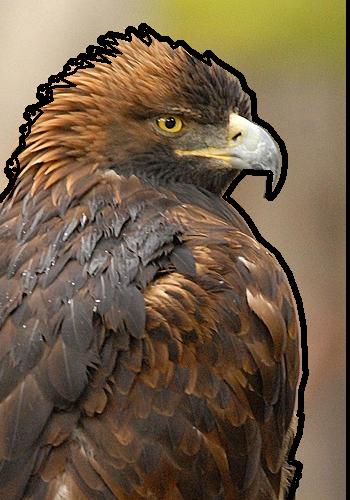}}
 \caption{Example of images in the training set and with mask I and II. }
\label{fig:example_VOC}
\end{figure}

\begin{itemize}
\item  \textbf{Mask I}: to mimic the holes in dis-occluded regions, which is generally around foreground objects' boundaries. The mask is designed as the dilated object boundaries. An example is shown in Fig.\ref{fig:maskI}.

\item  \textbf{Mask II}: to mimic the shifted objects' boundaries in the synthesized views induced by compression on depth map \cite{bosc2011towards}. We generate the second type of masks by simply shifting the first type of mask with certain pixels as shown in Fig.~\ref{fig:maskII}. 
\end{itemize}

\begin{figure}[!htbp]
 \mbox{ \parbox{0.5\textwidth}{
  \begin{minipage}[b]{0.15\textwidth}
   \subfloat[Supper-pixel map]
  {\label{fig:super_pixel_a}\includegraphics[width=\textwidth]{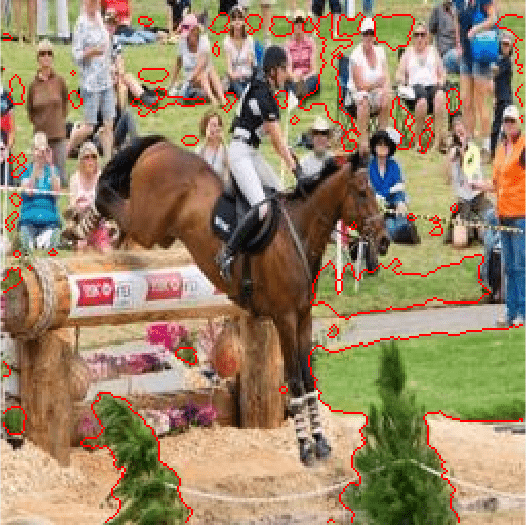}}  
  \end{minipage}
    \begin{minipage}[b]{0.15\textwidth}
  \subfloat[Small size mask]
  {\label{fig:small_mask_a}\includegraphics[width=\textwidth]{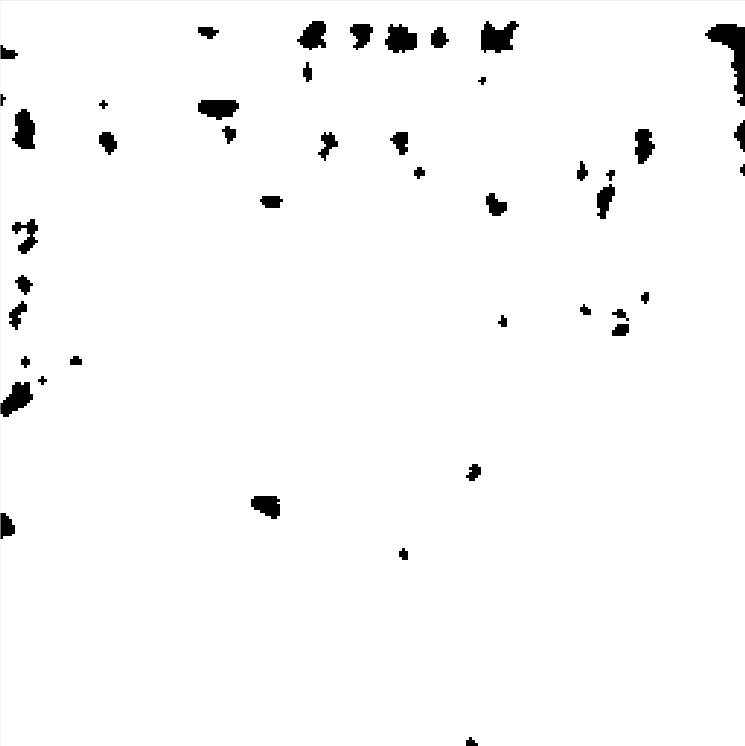}}
  \end{minipage}
    \begin{minipage}[b]{0.15\textwidth}
    \subfloat[Medium size mask   ]
  {\label{fig:medium_mask_a}\includegraphics[width=\textwidth]{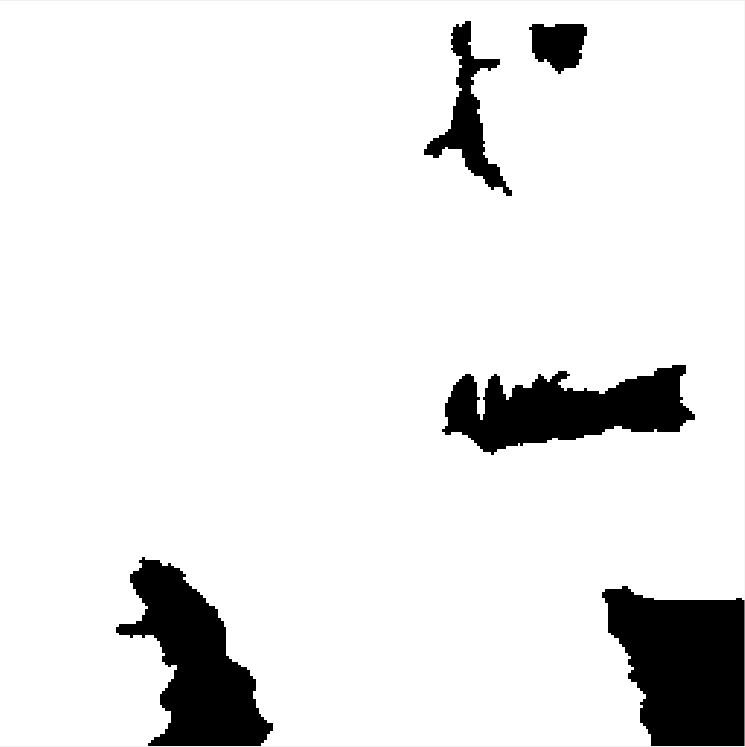}}
   \end{minipage} 
  \\
    \begin{minipage}[b]{0.15\textwidth}
    \subfloat[Supper-pixel map]
  {\label{fig:super_pixel_b}\includegraphics[width=\textwidth]{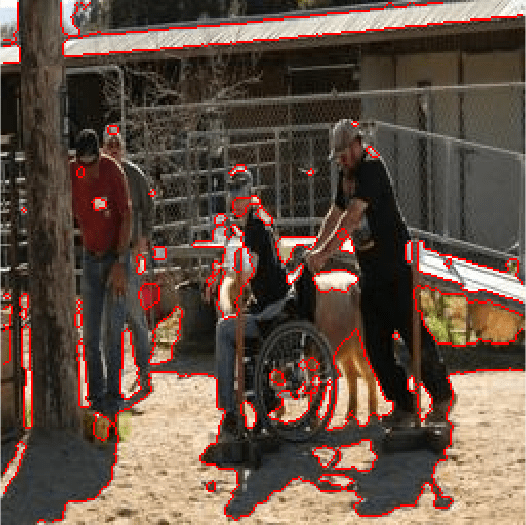}}
    \end{minipage}
    \begin{minipage}[b]{0.15\textwidth}
    \subfloat[Small size mask]
  {\label{fig:small_mask_b}\includegraphics[width=\textwidth]{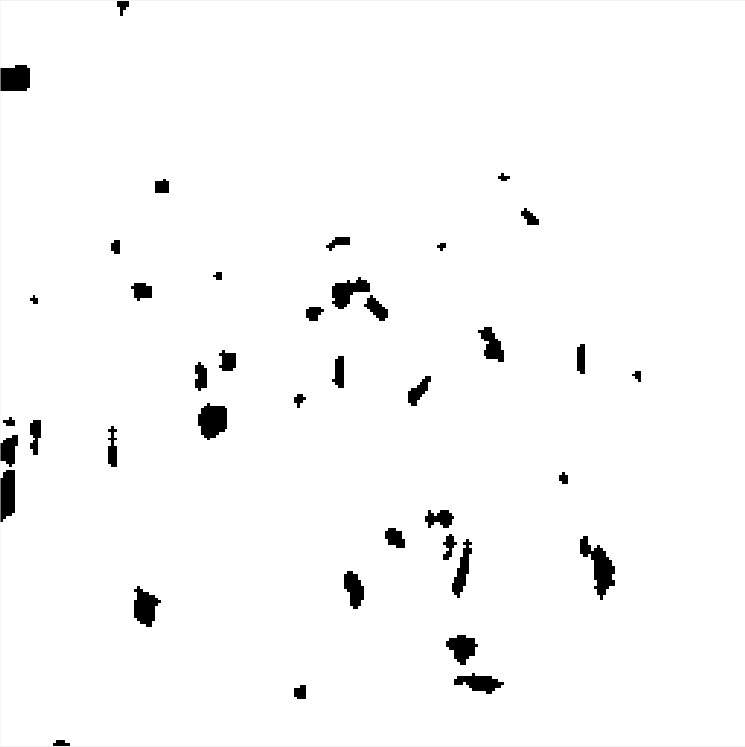}}
    \end{minipage}
    \begin{minipage}[b]{0.15\textwidth}
    \subfloat[Medium size mask]
  {\label{fig:medium_mask_b}\includegraphics[width=\textwidth]{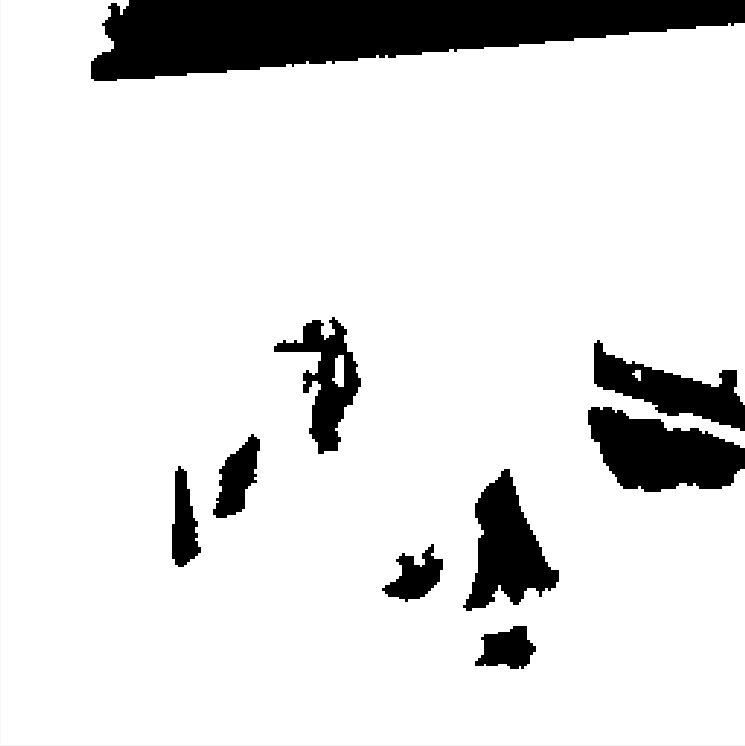}} 
\end{minipage}
}}
\caption{Example of images in the training set and designed mask III. Two mask sizes are considered.}
\label{fig:example_scene}
\end{figure}

Generally, it is easier to inpaint smooth regions with homogeneous textures than complicated regions with non-homogeneous textures, as the context around a smoother region is more `copyable' and less structure are involved. Hence, the quality of smooth inpainted regions within homogeneous texture is generally better than the non-homogeneous ones. If one wants to train a more powerful context encoder, the selected masks should contain contents/structures that can not be replicated from the surroundings. In addition, unlike the big connected region masks with arbitrary location introduced in \cite{pathak2016context}, dis-occluded regions or missing parts in a virtual view are generally disconnected, and the shapes of these regions are always related to the foreground objects (i.e., related to the depth map). With these two concerns, the third mask is proposed:

\begin{itemize}
\item \textbf{Mask III}: The SLIC super-pixels algorithm \cite{achanta2010slic} is used to select regions where masks should be located for later training. More specifically, an image is first segmented into a set of super-pixels as shown in Fig. \ref{fig:super_pixel_a}  and \ref{fig:super_pixel_b}. Then, \textbf{two mask sizes are considered}. Super-pixels that contains less than 100 pixels are considered as small size mask, while super-pixel contains 200 to 1000 pixels are considered as medium size mask. Examples are presented in Fig. \ref{fig:example_scene}. The black holes in Fig. \ref{fig:small_mask_a}  and \ref{fig:small_mask_b} are small size masks which are similar to the small holes shown in Fig. \ref{fig:example_holes1}. The holes in Fig. \ref{fig:medium_mask_a} and \ref{fig:medium_mask_b} are with medium size. By doing so, 1) the selected masks are separately distributed in the entire image; 2) the shape of the masks are related to objects; 3) the content within each mask region is more independent from its neighborhood.
\end{itemize}

\space


\subsection{Bag-of-Distortion-Words (BDW) codebook learning with pre-trained discriminator}
 \label{sec:BQW}
 
As discussed before, the discriminator serves as an indicator telling whether a patch is well inpainted or not. Thus the output of the discriminator is related to the quality of the patch. Therefore, it is reasonable to hypothesize that the intermediate output of $D$ is strongly related to inpainting related distortions which affect the perceived quality significantly. Based on this hypothesis, we propose to use the discriminator to get a latent codebook with `codewords' that represent different types of distortions. With this codebook, a higher-level representation could be obtained for each image. Details are illustrated below.

To predict the image level quality by considering local distortion, the image needed to be processed locally. Therefore, a set of multiple overlapping patches $P_i=\{p_{ij} | j=1,\cdots,n \}$, where $n$ is the total number of patches, is used to represent the image $x_i$ as done in \cite{li2017exploiting}. In this study, the overlap size is selected as half of the patch size, and the patches are sampled over the whole image  (along both the horizontal and vertical direction) to maintain as much structural information as possible. Afterwards, with the pre-trained GANs model, these patches are fed into the adversarial discriminator to extract higher-level features for later patches categorizations. For each patch $p_{ij}$ in the entire dataset $I$ for codebook training, its corresponding feature vector $v_{ij}$ is extracted from the $l_{th}$ layer in the discriminator as: 

\begin{equation}
v_{ij} = D(p_{ij},l) 
\end{equation}
In this study, the feature vector is extracted from the last convolutional layer of the discriminator (Details are shown in Section \ref{subsec:local_dis_region_selection}). Finally, $m \times n$ vectors can be obtained for the $m$ images in the codebook training set $I$. 

\begin{figure}[!htbp]
 \mbox{ \parbox{0.5\textwidth}{
  \begin{minipage}[b]{0.15\textwidth}
   \subfloat[$c_{47}$]
  {\label{fig:figB}\includegraphics[width=\textwidth]{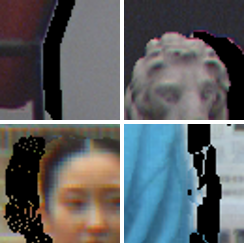}} 
  \end{minipage}
    \hspace{0.1cm}
    \begin{minipage}[b]{0.15\textwidth}
  \subfloat[$c_{147}$]
  {\label{fig:figA}\includegraphics[width=\textwidth]{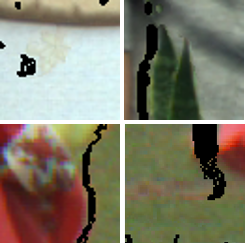}}
  \end{minipage}
    \hspace{0.1cm}
    \begin{minipage}[b]{0.15\textwidth}
    \subfloat[$c_{102}$   ]
  {\label{fig:figC}\includegraphics[width=\textwidth]{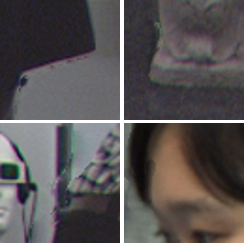}}
   \end{minipage} 
  \\
    \begin{minipage}[b]{0.15\textwidth}
    \subfloat[$c_{121}$]
  {\label{fig:figC}\includegraphics[width=\textwidth]{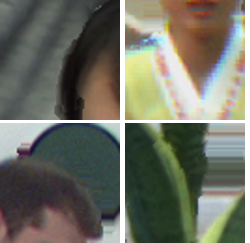}}
    \end{minipage}
      \hspace{0.1cm}
    \begin{minipage}[b]{0.15\textwidth}
    \subfloat[$c_{05}$]
  {\label{fig:figC}\includegraphics[width=\textwidth]{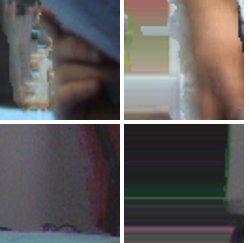}}
    \end{minipage}
      \hspace{0.1cm}
    \begin{minipage}[b]{0.15\textwidth}
    \subfloat[$c_{98}$]
  {\label{fig:figC}\includegraphics[width=\textwidth]{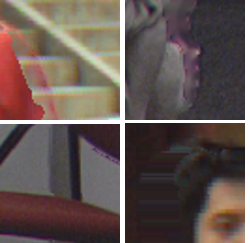}} 
\end{minipage}
}}
\caption{Selected `Words' in the learned BDW Codebook.}
\label{fig:cluster}
\end{figure}

With the set of extracted features in correspondence to their patches, now we want to look for a new representation of the entire image by taking the intermediate output of the discriminator into account so that this new representation is able to link the local information with the entire image quality.  

Intuitively, the idea is to categorize image patches into different clusters that can be representatives of perceived quality, so the quality of the tested image can be quantified by checking how many `good' or `poor' patches it has. Formally, the $m \times n$ patches $v_{ij}$, $i=\{1,\cdots,m\}$, $j=\{1,\cdots,n\}$ are reshaped to $v_o,o=\{1, \cdots ,n \times m\}$. Then the $v$ are clustered into $K$ clusters $\{ c_1, \cdots, c_K  \}$ using an advanced clustering algorithm~\cite{muja2014scalable}, which is a fast nearest neighbor algorithm robust in high dimensional vectors matching. Selected cluster results are shown in Fig. \ref{fig:cluster}. It can be observed that patches with similar type of distortions are gathered in the same cluster as a `distortion word'. For example, both of the cluster $c_{47}$ and $c_{147}$ are consist of patches with `dark holes', and the ones in cluster $c_{47}$ are obviously larger than that of $c_{147}$, which indicates worse quality. For other clusters shown in the figure, the distortions of $c_{102}$  is imperceivable (guarantee good quality), while $c_{121}$, $c_{05}$ and $c_{98}$ are with more obvious inpainting related artifacts. Naturally, different `codeword' in the clustered `codebook' actually represents a certain level of quality with respect to the types and magnitudes of distortions, which is in consistent with our hypothesis. Based on this observation, in this study, the trained codebook is named after `bag of distortion words' (BDW). With the BDW codebook, each image $x_i$ can then be encoded as a histogram $h_{adv}(i)= \{ \mu_{i1}, \cdots, \mu_{iK}  \}$ , where each $\mu_{ik}$ is defined as

\begin{equation}
\mu_{ik} = \frac{ \sum_{j=1}^{n_p} \textbf{1}(p_{ij} \subset  c_k)}{n_p}
\label{eq:freq}
\end{equation}

\noindent $\textbf{1}(c)$ is an indicator function that equals to 1 if the specified binary clause $c$ is true, and $n_p$ is the number of patches within the image. An intuitive interpretation of this BDW based representation of the image is that the histogram statistically quantifies how many `good quality' and `poor quality' patches that a synthesized image has. As local significant synthesized distortion is more annoying than the global uniform one, this new representation is a higher-level quality descriptor which can indirectly predict the overall quality of one image. During clustering, $K$ is an important parameter that will have an impact on the final performance. Therefore, further discussion about the selection of $K$ is given in Section \ref{subsec:K}.      

\subsection{Local distortion regions selection}
\label{subsec:local_dis_region_selection}

Generally, artifacts located at a region of interest is much more annoying than those located at an inconspicuous area~\cite{ninassi2007does}. In our case, `poor' quality regions (\textit{i.e.}, holes and inpainting artifacts) are generally in the regions of interest (such as the foreground object), thus, they are more likely to attract observers than the `good' ones. Therefore, images with even a small number of `poor' regions are penalized more gravely by the observers. Accordingly, it is reasonable to do the same penalization in the objective model as well.

Moreover, as discussed before, the discriminator is trained to distinguish artificial generated picture (inpainted images in this case) from the real one. A well-trained discriminator is supposed to be able to indicate poor inpainted regions. The output of the discriminator $D$ is a boolean value indicating whether the input patch $p_{ij}$ is an inpainted or not, where `1' for real patches and `0' for generated patches.  It is intuitive to hypothesize that patches assigned with `0' by the discriminator are those with poor quality. Hence, the discriminator is further utilized as a `poor' quality patches selector. As thus, Equation \eqref{eq:freq} could be modified to: 
 \begin{equation}
 \mu_{ik} = \frac{ \sum_{j=1}^{n_p} \textbf{1}(p_{ij} \subset  c_k) \cdot XOR( D(p_{ij}),1) )}{n_p}
\label{eq:improved}
\end{equation}
where $D( \cdot )$ is the direct boolean output of the pre-trained discriminator when taking a patch $p_{ij}$ as the input. $XOR(\cdot)$ is the exclusive OR operation, $ XOR( D(p_{ij}),1) )$ equals to 1 if $D(p_{ij})=0$. 

Apart from using the final boolean output of the discriminator for selecting the possible inpainted regions, another possibility is to use the output just before the final sigmoid layer (\textit{i.e.}, the last convolutional layer) with normalization. To do this, the output of the last convolutional layer of the discriminator for all the training patches $p_{ij}, i=\{1,\cdots,m\}, j=\{1,\cdots,n\}$ are collected and normalized into a range of $[0,1]$. After the normalization, the output of the last convolutional layer serves as a probability value indicating that whether the test patch is natural (non-inpainted) or not. A smaller value represents a higher probability that this patch is inpainted and with a greater magnitude of distortions. Afterwards, patches that with a certain magnitude of in-painting distortions can be selected according to a certain threshold $\varepsilon$, meaning that only poorly inpainted regions with certain low-quality level are selected for the final quality decision. By doing so, Equation~\eqref{eq:improved} could be further rewritten as:
\begin{equation}
 \mu_{ik} = \frac{ \sum_{j=1}^{n_p} \textbf{1}(p_{ij} \subset  c_k) \cdot \textbf{1}( D_{BS}(p_{ij}) <\varepsilon )}{n_p}
\label{eq:improved_threshold}
\end{equation}
where $D_{BS}(\cdot) $ means that we only consider the output of the last convolutional layer in the discriminator with a patch $p_{ij}$ as input. $\varepsilon$ is a threshold for poor-quality patches selection. The setting of threshold $\varepsilon$ is discussed in Section \ref{subsec:threshold}.

\subsection{Final quality prediction}
\label{sec:prediction}
After extracting the histogram $h_{adv}$,  Support Vector Regression (SVR) is then applied on it with a linear kernel to predict the final quality score. In the experiment, the entire database is divided into 20\% validation set for model parameters selection (\textit{e.g.}, codebook training) and 80\% for performance evaluation. During the performance evaluation procedure, a 1000-fold cross-validation is applied. For each fold, the remaining 80\% of the dataset is further randomly split into 80\% of the images for SVR training and 20\% for testing, with no overlap between them \cite{gastaldo2013supporting} (no same viewpoint of the same content). The median Pearsons Correlation Coefficient (PCC), Spearman rank order Correlation Coefficient (SCC), and Root Mean Square Error (RMSE) between subjective and objective scores are reported across the 1000 runs for performance evaluation. 
 
\section{Experimental Result}
\label{sec:exp}
The performance of the proposed GANs-NQM is evaluated on IRCCyN/IVC DIBR images database~\cite{bosc2011towards}. Images from this database are obtained from three multi-view RGB-D sequences: Book Arrival, Lovebird1, and Newspaper. Seven RGB-D synthesis algorithms labeled with A1-A7~\cite{fehn2004depth,telea2004image,mori2009view,mueller2009view,ndjiki2011depth,koppel2010temporally} are used to process the three sequences to generate four new virtual views for each of them. The database is composed of 84 synthesized views and 12 original frames extracted from the corresponding sequences along with subjective quality scores in the form of mean opinion score (MOS). 

In our study, as the objective is to evaluate the quality of RGB-D synthesis views, differential MOS (DMOS)~\cite{itu2014subjective} are calculated and used. Data augmentation is conducted to provide more robust performance evaluation by rotating each image in the database $90^\circ$, $180^\circ$ and $270^\circ$ counterclockwise successively, which ends up into totally 384 images. Unlike other data augmentation methodology (such as scaling), rotation operation does not introduce new distortion. We thus assume that the qualities of the augmented images remain unchanged. The performance evaluation procedure is conducted  according to~\cite{gastaldo2013supporting} as described in section~\ref{sec:prediction}. For training the BDW codebook, 20 \% of the augmented data is utilized as validation set, which contains around  $1.5 \times 10^4$ pathces of size $64 \times 64$ . 

\subsection{  GANs based dis-occluded regions inpainter }
\subsubsection{ Training data }
 To generate a new dataset with the three masks mentioned above, we collect images from the PASCAL VOC 2012 \cite{everingham2015pascal} and the Places database \cite{zhou2017places}. There are in total 10K training images in this study.

\begin{itemize}  
\item{\textbf{PASCAL VOC 2012 database}}: The original objective of this database is for a challenge to recognize objects from a number of visual object classes in realistic scenes. It contains 3K images with twenty object classes, which diverse from people, animals to vehicles and indoor scenes. One of the merits of this database is that it provides us with pixel-wise segmentation labels, which gives the boundary of `objects' against the 'background' label. An example is given in Fig. \ref{fig:original_img} and Fig. \ref{fig:labeled_map}. In our study, we utilize this segmentation label to generate Mask I and Mask II mentioned above, which leads to 6K training data. 
 
\item{\textbf{Places database}}: To have a balanced dataset with mask I and II, the validation set from the `Places Challenge 2017', which contains around 2K images, are selected as a part of the training set in this study with mask III mentioned above. This dataset contains images with diverse contents, which vary from outdoor landscapes, cities views to indoor people portrait images. As there are two mask sizes in Mask III,  this leads to 4K training images.
  
\end{itemize}
 
\subsubsection{ Training process }
The framework of the `context inpainter' is implemented based on the pipeline developed by Pathak~\textit{et al.}~\cite{pathak2016context} with Caffe and Torch packages. The commonly used stochastic gradient descent method Adam\cite{kinga2015method} is used for optimization. We start with a learning rate of 0.0002, as set in DCGAN~\cite{radford2015unsupervised}, but a different bottleneck of 4000 units. In our experiment, the impact of trade-off between \textit{G} and \textit{D}, ~\textit{i.e.}, different $\lambda$ in Equation (\ref{eq:l_joint}),  on the performance of the proposed metric has been tested.   

For the architecture of the GANs network, as it has been tested in \cite{pathak2016context} that finer inpainting results can be obtained by replacing pooling layers with the convolutional ones, in this study, the pool-free structure remains. Furthermore, since the main focus of this section is to explore the discriminator for quality assessment of synthesized views with local non-uniform distortions, we only change the architecture of the discriminator. Details of all the discriminator architectures that have been tested in this study are summarized in Table \ref{tab:D_A}. The main difference between $D_1$ and the other two architectures is the size of images that can be fed into. $D_2$ is of less complex structures than $D_3$ and $D_1$, where the number of convolutional kernels is halved in each layer. With such design, we could check how the input size and complexity of the discriminator influence the performance of the proposed scheme.

\begin{table*}[!htb]
\centering
\caption{Different discriminator architectures tested in this study, \textbf{$In$} is the input of each layer, ~\textbf{$InSize$} is the input size of each layer,~\textbf{$k$} is the kernel size,~\textbf{s} is the stride, ~\textbf{$OutL$} is the output channels for each layer and \text{$Act$} is the activation function of each layer.}
\begin{tabular}{|c|c|c|c|c|c|c|c|m{10cm}|}\hline
Layer &  In  &  InSize & k  & s & OutL & Act  & Visualization \\  \hline
\multicolumn{8}{ |c| }{ Discriminator architecture $D_1$ }\\ \hline
\multirow{2}{*}{conv\_1 }& \multirow{2}{*}{image} &  \multirow{2}{*}{$64 \times 64$} & \multirow{2}{*}{4}  &  \multirow{2}{*}{1} & \multirow{2}{*}{64} & Leaky & \\  
  &  &  &   &   &   &ReLU & \multirow{4}{*}{\includegraphics[ width=.4\textwidth]{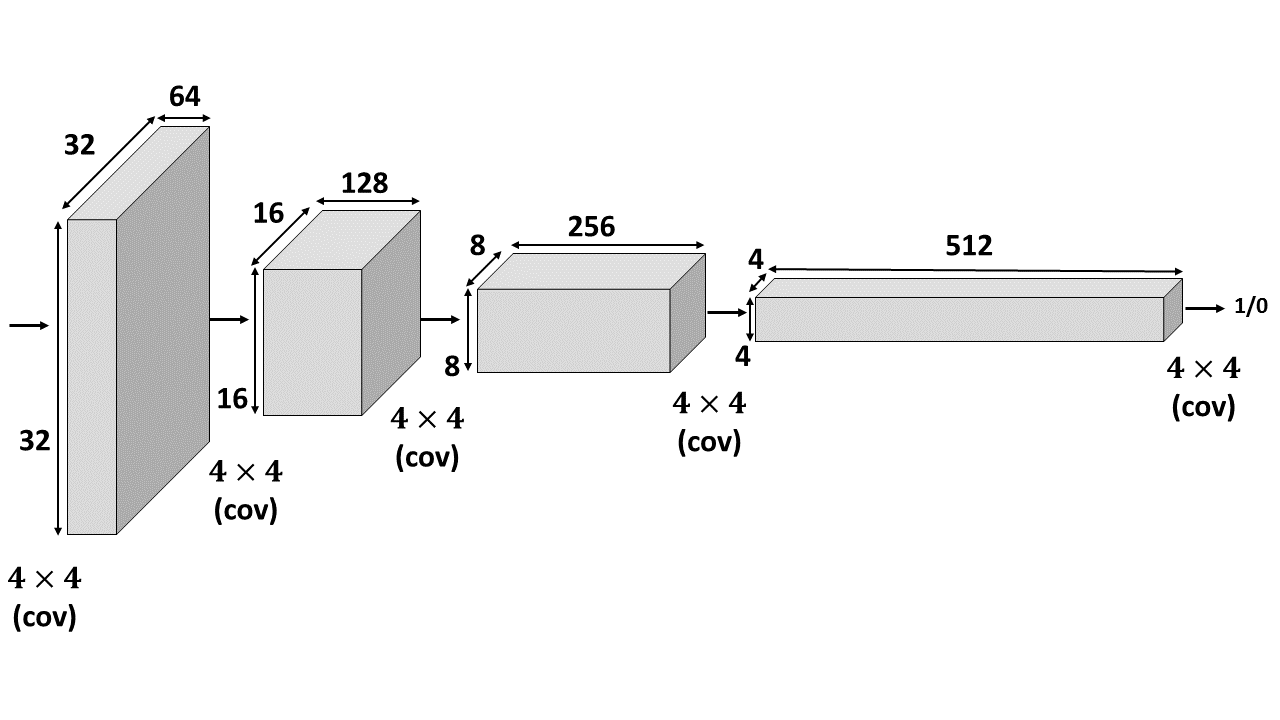}}   \\ \cline{1-7}  
  
\multirow{2}{*}{conv\_2 }& \multirow{2}{*}{conv\_1} &  \multirow{2}{*}{$32 \times 32$} & \multirow{2}{*}{4}  &  \multirow{2}{*}{1} & \multirow{2}{*}{128} & Leaky &  \\  
  &  &  &   &   &   &  ReLU &   \\ \cline{1-7}

\multirow{2}{*}{conv\_3 }& \multirow{2}{*}{conv\_2} &  \multirow{2}{*}{$16 \times 16$} & \multirow{2}{*}{4}  &  \multirow{2}{*}{1} & \multirow{2}{*}{256} &  Leaky &  \\   
 &  &  &   &   &   & ReLU  &  \\ \cline{1-7}

\multirow{2}{*}{conv\_4 }& \multirow{2}{*}{conv\_3} &  \multirow{2}{*}{$8 \times 8$} & \multirow{2}{*}{4}  &  \multirow{2}{*}{1} & \multirow{2}{*}{512} & Leaky &  \\  
  &  &  &   &   &   &  ReLU  & \\ \cline{1-7}  
  
 \multirow{2}{*}{conv\_5 }& \multirow{2}{*}{conv\_4} &  \multirow{2}{*}{$4 \times 4$} & \multirow{2}{*}{4}  &  \multirow{2}{*}{1} & \multirow{2}{*}{1} &  Sig &  \\  
  &  &  &   &   &   & moid &  \\  \hline
  \multicolumn{8}{ |c| }{ Discriminator architecture $D_2$ }\\ \hline
\multirow{2}{*}{conv\_1 }& \multirow{2}{*}{image} &  \multirow{2}{*}{$128 \times 128$} & \multirow{2}{*}{4}  &  \multirow{2}{*}{1} & \multirow{2}{*}{32} & Leaky & \\  
  &  &  &   &   &   &ReLU & \multirow{4}{*}{\includegraphics[ width=.4\textwidth]{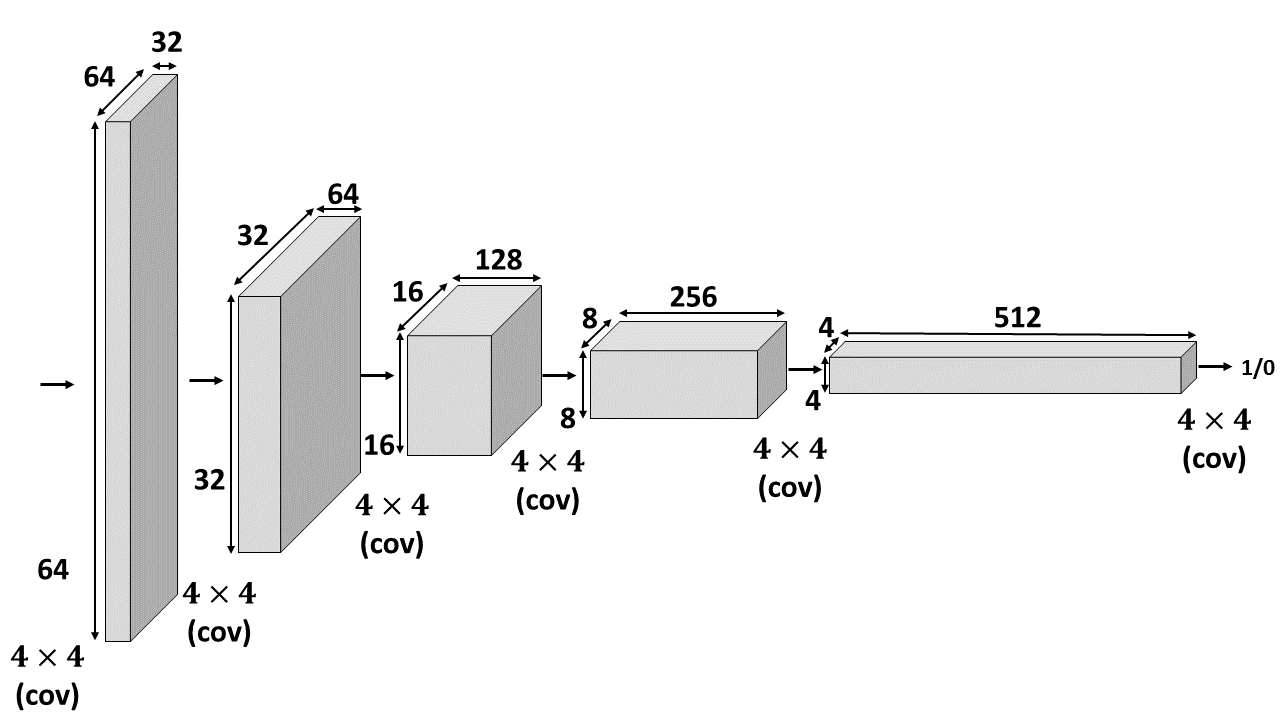}}  \\ \cline{1-7}  
  
\multirow{2}{*}{conv\_2 }& \multirow{2}{*}{conv\_1} &  \multirow{2}{*}{$64 \times 64$} & \multirow{2}{*}{4}  &  \multirow{2}{*}{1} & \multirow{2}{*}{64} & Leaky &  \\  
  &  &  &   &   &   &  ReLU &  \\ \cline{1-7}

\multirow{2}{*}{conv\_3 }& \multirow{2}{*}{conv\_2} &  \multirow{2}{*}{$32 \times 32$} & \multirow{2}{*}{4}  &  \multirow{2}{*}{1} & \multirow{2}{*}{128} &  Leaky &  \\  
  &  &  &   &   &   & ReLU  &  \\ \cline{1-7}

\multirow{2}{*}{conv\_4 }& \multirow{2}{*}{conv\_3} &  \multirow{2}{*}{$16 \times 16$} & \multirow{2}{*}{4}  &  \multirow{2}{*}{1} & \multirow{2}{*}{256} & Leaky &  \\  
  &  &  &   &   &   &  ReLU  & \\ \cline{1-7}  
  
 \multirow{2}{*}{conv\_5 }& \multirow{2}{*}{conv\_4} &  \multirow{2}{*}{$8 \times 8$} & \multirow{2}{*}{4}  &  \multirow{2}{*}{1} & \multirow{2}{*}{512} &   Leaky &  \\  
  &  &  &   &   &   & ReLU  &  \\  \cline{1-7}     
\multirow{2}{*}{ conv\_6 }& \multirow{2}{*}{ conv\_5 } &  \multirow{2}{*}{ $4 \times 4$ } & \multirow{2}{*}{ 4 }  &  \multirow{2}{*}{ 1 } & \multirow{2}{*}{ 1 } & Sig&  \\  
  &  &  & &   &   & moid &  \\ \hline
  
  \multicolumn{8}{ |c| }{ Discriminator architecture $D_3$ }\\ \hline
\multirow{2}{*}{conv\_1 }& \multirow{2}{*}{image} &  \multirow{2}{*}{$128 \times 128$} & \multirow{2}{*}{4}  &  \multirow{2}{*}{1} & \multirow{2}{*}{16} & Leaky &\\  
  &  &  &   &   &   &ReLU &  \multirow{4}{*}{\includegraphics[ width=.4\textwidth]{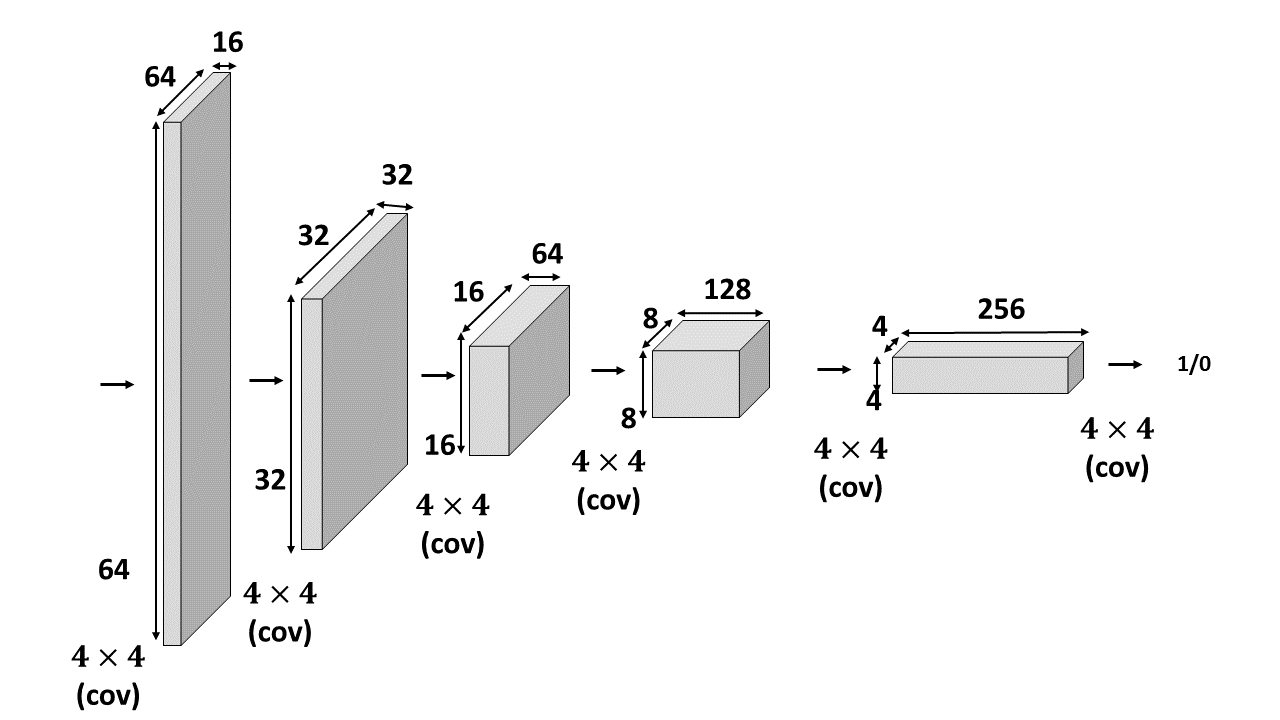}}  \\ \cline{1-7}  
  
\multirow{2}{*}{conv\_2 }& \multirow{2}{*}{conv\_1} &  \multirow{2}{*}{$64 \times 64$} & \multirow{2}{*}{4}  &  \multirow{2}{*}{1} & \multirow{2}{*}{32} & Leaky &  \\  
  &  &  &   &   &   &  ReLU &  \\ \cline{1-7}

\multirow{2}{*}{conv\_3 }& \multirow{2}{*}{conv\_2} &  \multirow{2}{*}{$32 \times 32$} & \multirow{2}{*}{4}  &  \multirow{2}{*}{1} & \multirow{2}{*}{64} &  Leaky &  \\  
  &  &  &   &   &   & ReLU  &  \\ \cline{1-7}

\multirow{2}{*}{conv\_4 }& \multirow{2}{*}{conv\_3} &  \multirow{2}{*}{$16 \times 16$} & \multirow{2}{*}{4}  &  \multirow{2}{*}{1} & \multirow{2}{*}{128} & Leaky &  \\  
  &  &  &   &   &   &  ReLU  & \\ \cline{1-7}  
  
 \multirow{2}{*}{conv\_5 }& \multirow{2}{*}{conv\_4} &  \multirow{2}{*}{$8 \times 8$} & \multirow{2}{*}{4}  &  \multirow{2}{*}{1} & \multirow{2}{*}{256} &   Leaky &  \\  
  &  &  &   &   &   & ReLU  &  \\  \cline{1-7}     
\multirow{2}{*}{ conv\_6 }& \multirow{2}{*}{ conv\_5 } &  \multirow{2}{*}{ $4 \times 4$ } & \multirow{2}{*}{ 4 }  &  \multirow{2}{*}{ 1 } & \multirow{2}{*}{ 1 } & Sig&  \\  
  &  &  & &   &   & moid &  \\ \hline  
  
\end{tabular}
\label{tab:D_A}
\end{table*}

\subsection{Performance dependency of hyper parameters}
\subsubsection{Number of `Quality Word' $K$ in BQW}
\label{subsec:K}

\begin{figure}[!htbp]
  \includegraphics[width=1\columnwidth]{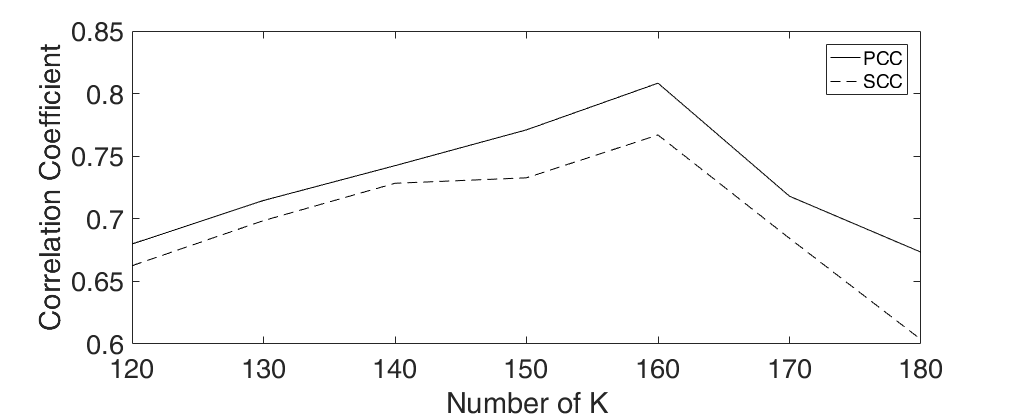}
  \caption{Performance dependency of proposed metric with changing $K$ number }
   \label{fig:K}
\end{figure}

To check if the performance of the proposed GAN-NQM is sensitive to the cluster number $K$, different numbers of $K$ for the quality-aware codebook training are tested on the validation set. The results are shown in Fig.~\ref{fig:K}. The corresponding PCC/SCC curves are obtained by fixing other related parameters. It can be observed that the performance of GAN-NQM in PCC/SCC raises gradually along with the increase of $K$ at the beginning. After the performance peaks at a certain number ($K = 160$), it starts to drop gradually. Thus, in this study, we set $K=160$.

\subsubsection{Solver hyper-parameter $\lambda$}
As introduced in \cite{pathak2016context,radford2015unsupervised}, the solver hyper-parameter $\lambda$ in equation (\ref{eq:l_joint}) is suggested to be set as 0.999. It is a tunable parameter balancing the reconstruction loss and the adversarial loss during training. Since the discriminator is utilized for both distortion regions selection, and higher level feature extraction in this study, higher weight for the adversarial loss is tested, \textit{i.e.} lower $\lambda$ in Equation (\ref{eq:l_joint}). The performances of the proposed model with different $\lambda$ are reported in Table \ref{tab:lambda}, where we fix $K=160$ and use the direct output of the discriminator for distortion region selection. By comparing the performances, interestingly, it is found that the PCC increases when $\lambda$ increases from 0.5 to 0.9, and drops when $\lambda=0.999$. In this study, we set $\lambda=0.9$.

\begin{table}[]
\centering
\caption{Performance dependency of proposed metric with different solver hyper-parameters $\lambda $}
\label{tab:lambda}
\begin{tabular}{|c|c|c|c|}\hline
PCC      & $\lambda=0.5$    & $\lambda=0.9 $ & $\lambda=0.999 $\\\hline  
$D_1$   &  0.7802 & \bf{0.8083} &  0.7821 \\ \hline
$D_2$   & 0.7377 & 0.7536 &  0.7339\\ \hline
$D_3$   & 0.7266 & 0.7280 &  0.7273 \\ \hline
\end{tabular}
\end{table}

\subsubsection{Different Discriminator architecture} 
The performances of the proposed model equipped with different discriminator architectures, which are described in Table~\ref{tab:D_A}, are also reported in Table \ref{tab:lambda}. It is found that, with any chosen $\lambda$, the proposed method always attains better PCC value with architecture $D_1$ than with $D_2$ or $D_3$. In the proposed model, we finally choose architecture $D_1$ for the discriminator.

\subsubsection{Threshold $\varepsilon$ }
\label{subsec:threshold}
The influence of the threshold $\varepsilon$ on the performance of GANs-NQM is illustrated in Table~\ref{tab:thres}. The performance of using a strategy of selecting a proper threshold in Equation \eqref{eq:improved_threshold} for distortion regions selections is better than using the direct output of the discriminator. The performance climbs with an increasing $\varepsilon$ until it reaches to 0.7, then the performance declines. In our model, we set $\varepsilon=0.7$.

\begin{table}[]
\centering
\caption{Performance dependency of proposed metric with different Threshold $\varepsilon$}
\label{tab:thres}
\begin{tabular}{|c|c|c|c|}\hline
      & \textbf{PCC} & \textbf{SCC} & \textbf{RMSE} \\ \hline
Boolean output of $D$ &   0.7463    &   0.7166    &  0.5045     \\\hline
$\varepsilon$ = 0.3 & 0.7525 & 0.7259  & 0.4995 \\\hline
$\varepsilon$ = 0.4 & 0.7631 & 0.7649   &  0.4593\\ \hline
$\varepsilon$ = 0.5 & 0.7889 & 0.7600   & 0.4546 \\\hline
$\varepsilon$ = 0.6 & 0.7963 & 0.7798   & 0.4176 \\\hline
$\varepsilon$ = 0.7 & \textbf{0.8195} & \textbf{0.7920 }  & \textbf{0.4016} \\\hline
$\varepsilon$ = 0.8 & 0.7704 & 0.7248   & 0.4723\\\hline
$\varepsilon$ = 1.0 &  0.7425 & 0.7062 &	 0.5023\\\hline

\end{tabular}
\end{table}

\subsection{Overall quality prediction performance}

The performance of the proposed metric is compared with all the quality metrics that are developed for assessing synthesized views' quality summarized in Section \ref{sec:3DM}. For fair comparisons, the median performances of the compared metrics are also reported under a 1000-fold cross-validation. 

Performance results are summarized in Table \ref{tab:main performance}. These metrics can be classified into FR or NR metrics. Parameters of GANs-NQM that yield the best performance are selected according to the previous discussion. It can be seen from Table \ref{tab:main performance} that our proposed method attains the best performance within the group of NR metrics in terms of PCC, SCC, and RMSE. The gain of GAN-NQM compared to the second best NR metric APT is 17\% in terms of PCC. Furthermore, even compared to FR metrics, its performance is comparable to the best performed metric ST-IQM.

\begin{table}[!htbp]
\begin{center}
\caption{\label{tab:main performance}%
Performance of the proposed metric and the state-of-the-art metrics }
{
\renewcommand{\baselinestretch}{1}\footnotesize
\begin{tabular}{|c|c|c|c|}
\hline

     &\bf{PCC} &\bf{SCC} &\bf{RMSE}      \\ \hline
\multicolumn{4}{ |c| }{ Full Reference Metrics (FR)}\\ \hline
3DSwIM \cite{battisti2015objective} & 0.7266  & 0.6421   & 0.4304 \\ \hline
VSQA \cite{conze2012objective} & 0.5096  & 0.5064   & 0.5336 \\ \hline
MP-PSNR$_{red}$ \cite{sandic2016dibr} &0.7489  & 0.7011  & 0.4148  \\ \hline
MP-PSNR$_{ful}$ \cite{sandic2015dibrMP} &0.7336  & 0.6634   & 0.4199  \\ \hline
MW-PSNR$_{red}$ \cite{sandic2016dibr}& 0.7400  & 0.6836  & 0.4240  \\ \hline
MW-PSNR$_{ful}$   \cite{sandic2015dibr} & 0.7183  & 0.6419  &  0.4401 \\ \hline
CT-IQM \cite{CT_IQA} &0.7107  & 0.6151   & 0.4481 \\ \hline
EM-IQM \cite{EM-IQA}& 0.7599  & 0.7012 & 0.4038 \\ \hline
ST-IQM \cite{ling2017image} & \bf{0.8462}  & \bf{0.7681}   & \bf{0.3415}  \\  \hline
\multicolumn{4}{ |c| }{NO Reference Metrics (NR)}\\ \hline
NIQSV+ \cite{tian2018niqsv+} & 0.7010  & 0.5158   & 0.4553 \\  \hline
APT \cite{gu2017model}&  0.7046  & 0.7198   & 0.4993  \\   \hline
GANs-NQM(proposed) &   \bf{0.8262}  &  \bf{0.8072}  & \bf{0.3861}   \\\hline

\end{tabular}}
\end{center}
\end{table}

\begin{figure*}[!htbp]

\subfloat[ ]{
\includegraphics[width=0.33\textwidth]{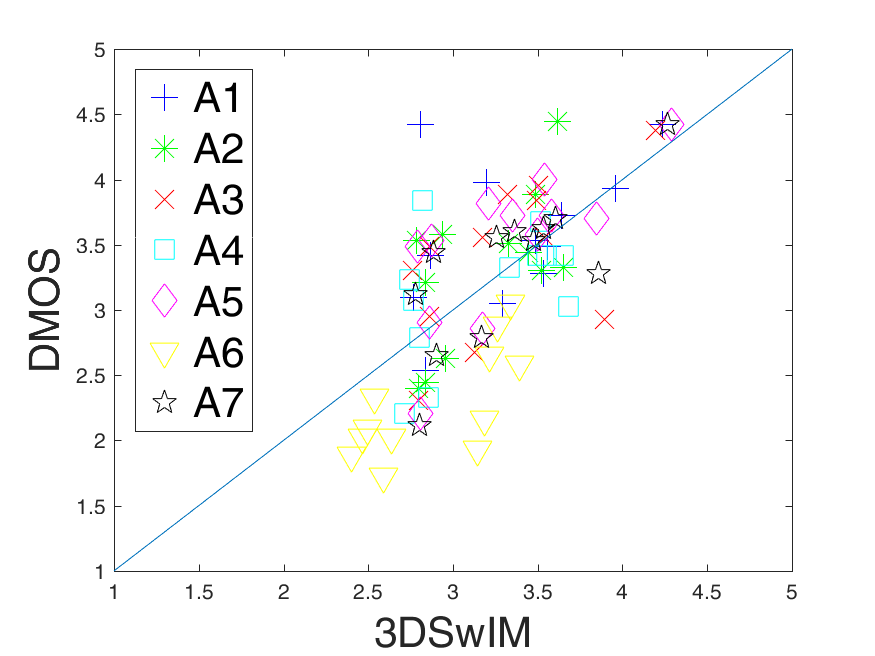}}
\subfloat[ ]{
\includegraphics[width=0.33\textwidth]{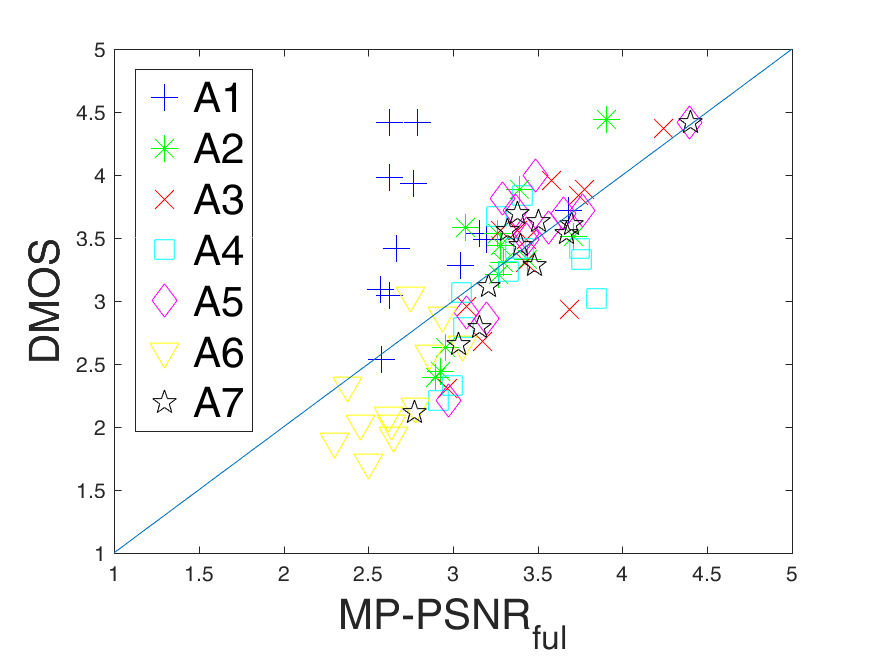}}
\subfloat[ ]{
\includegraphics[width=0.33\textwidth]{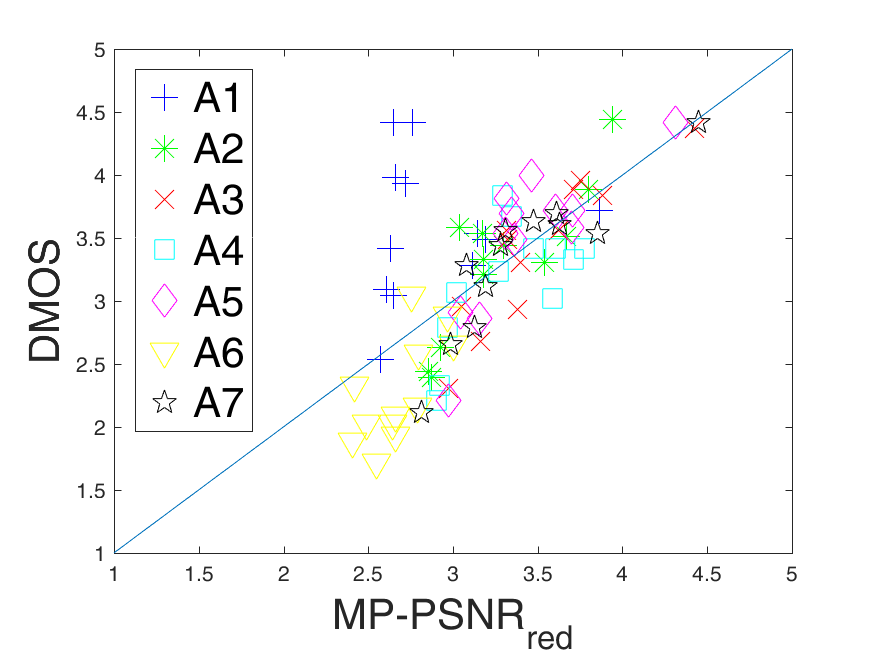}}
 \\ 

\subfloat[ ]{
\includegraphics[width=0.33\textwidth]{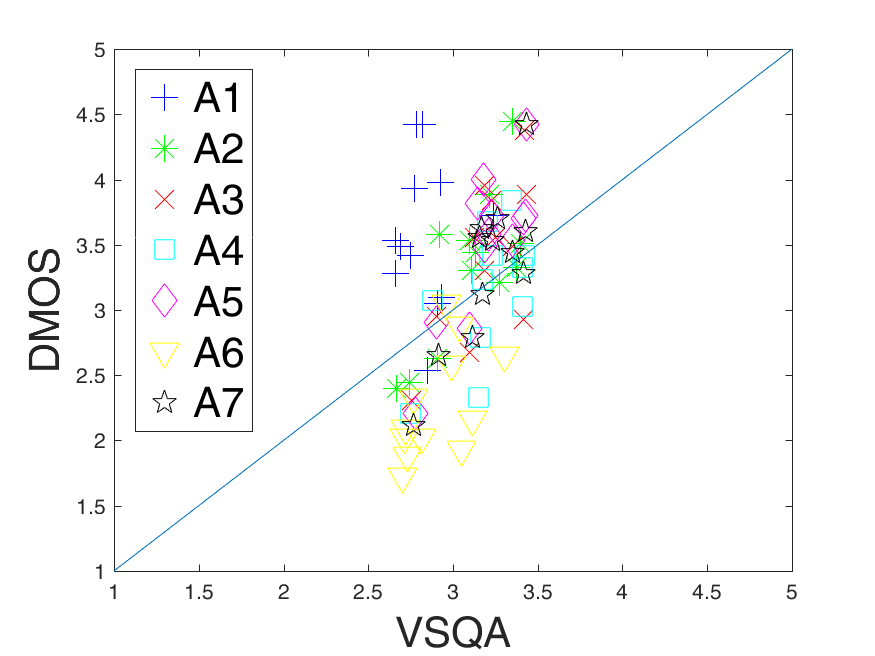}}
\subfloat[ ]{
\includegraphics[width=0.33\textwidth]{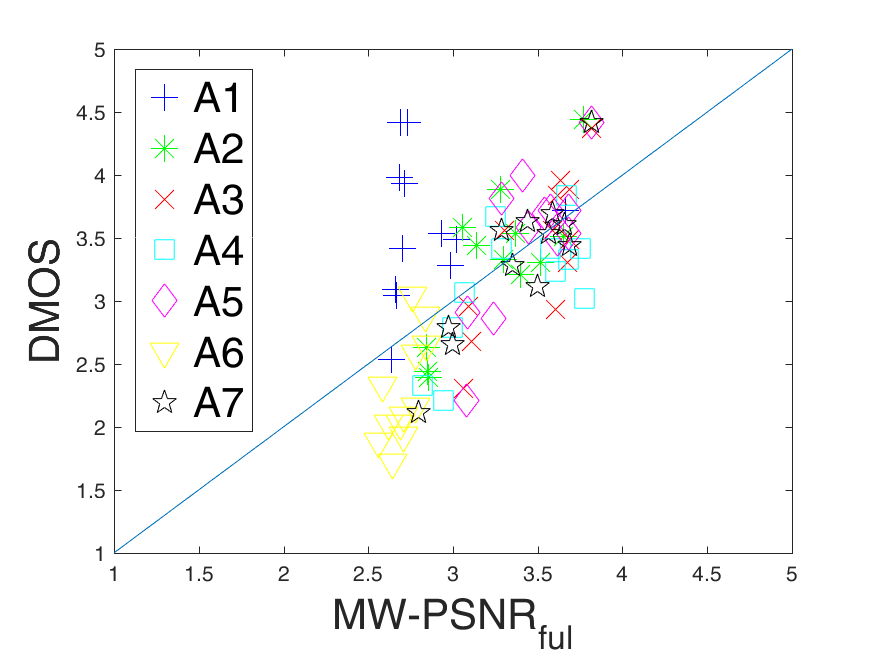}}
\subfloat[ ]{
\includegraphics[width=0.33\textwidth]{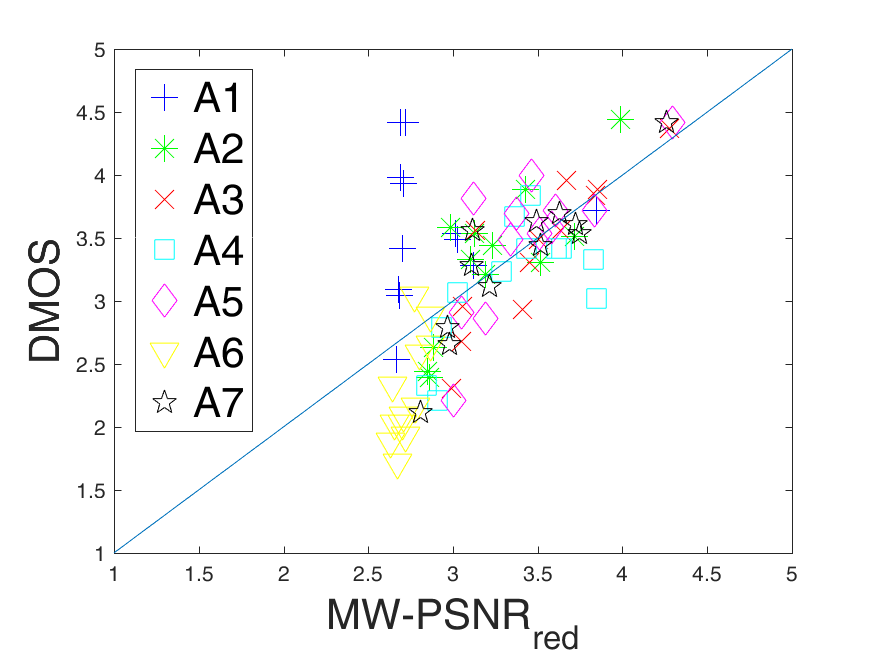}}
 \\ 

\subfloat[ ]{
\includegraphics[width=0.33\textwidth]{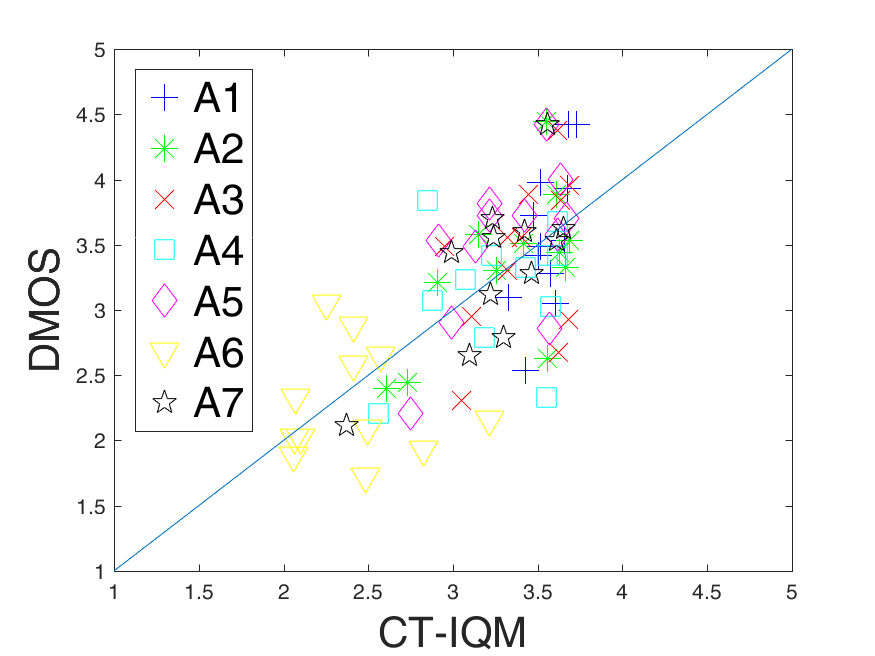}}
\subfloat[ ]{
\includegraphics[width=0.33\textwidth]{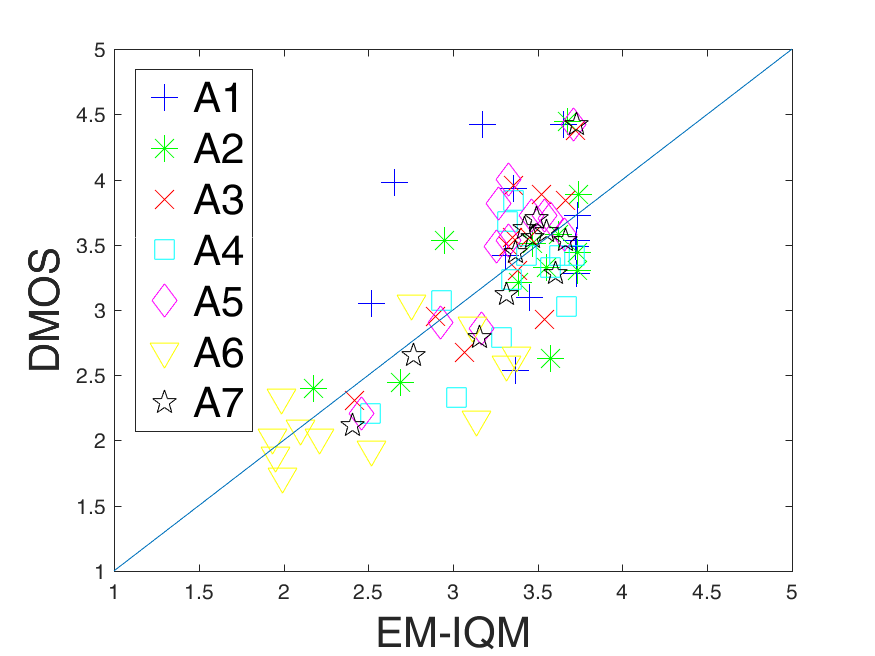}}
\subfloat[ ]{
\includegraphics[width=0.33\textwidth]{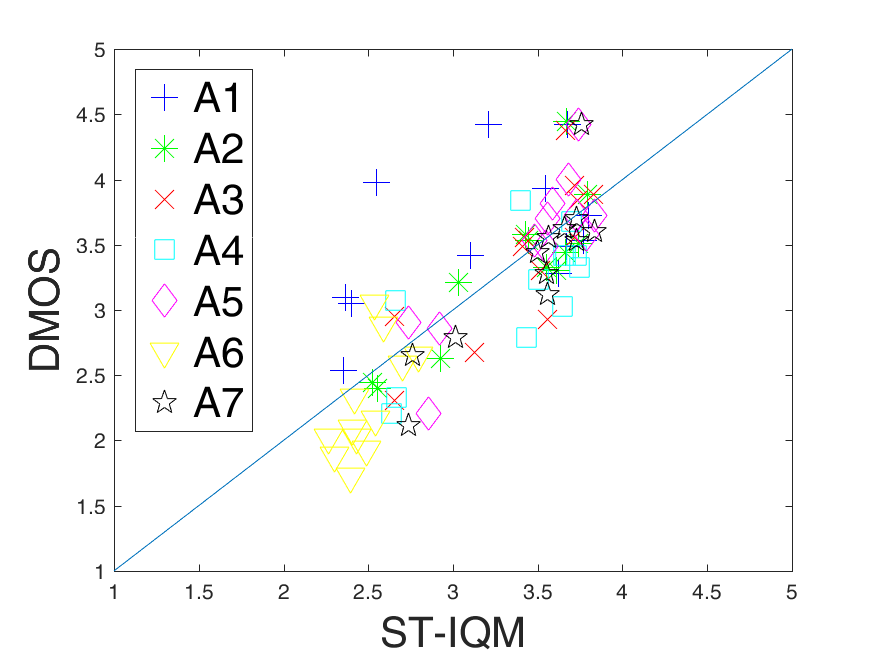}}
 \\ 

\subfloat[ ]{\label{fig:scatter_plot_apt}
\includegraphics[width=0.33\textwidth]{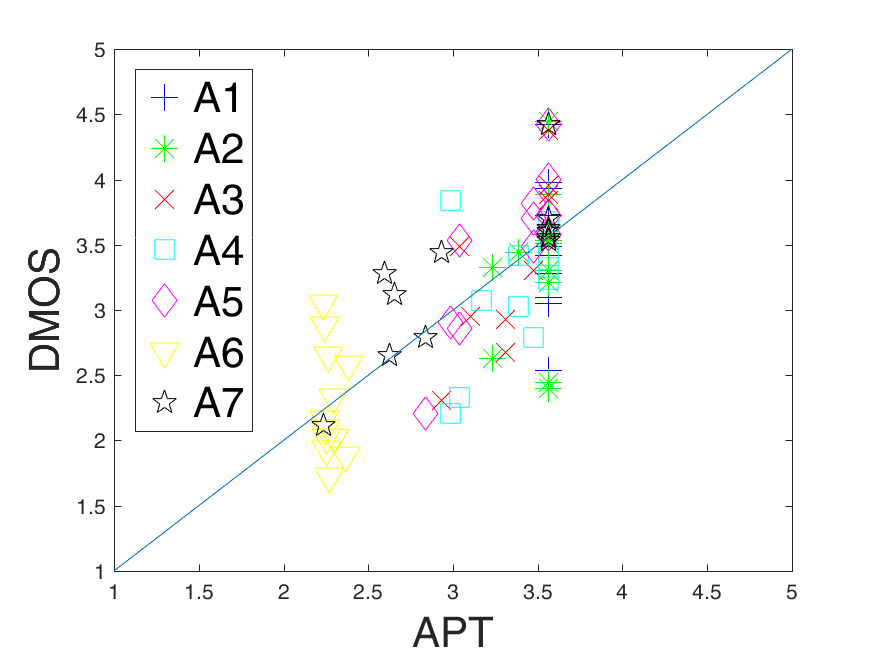}}
\subfloat[ ]{\label{fig:scatter_plot_niqsv+}
\includegraphics[width=0.33\textwidth]{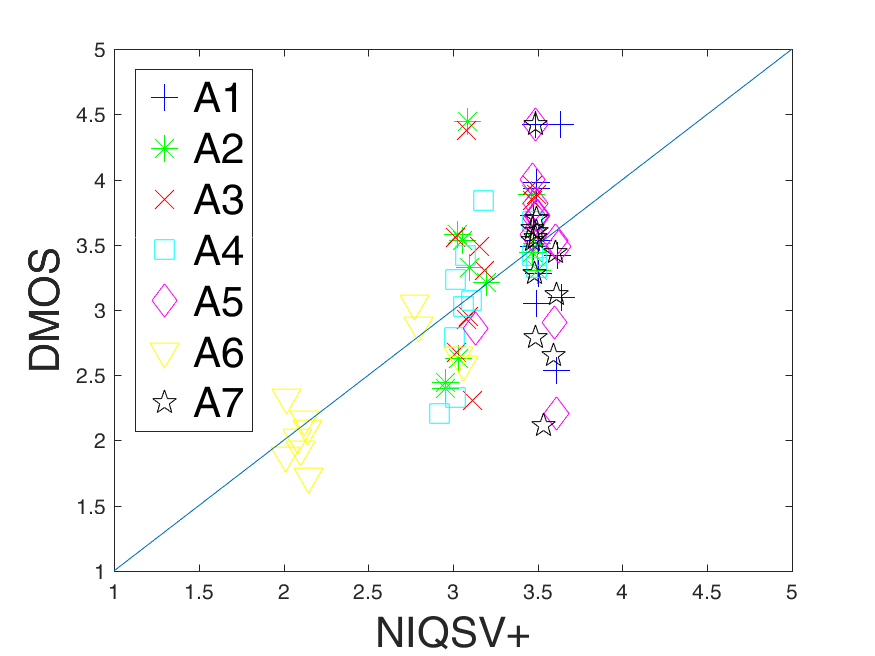}}
\subfloat[ ]{
\includegraphics[width=0.33\textwidth]{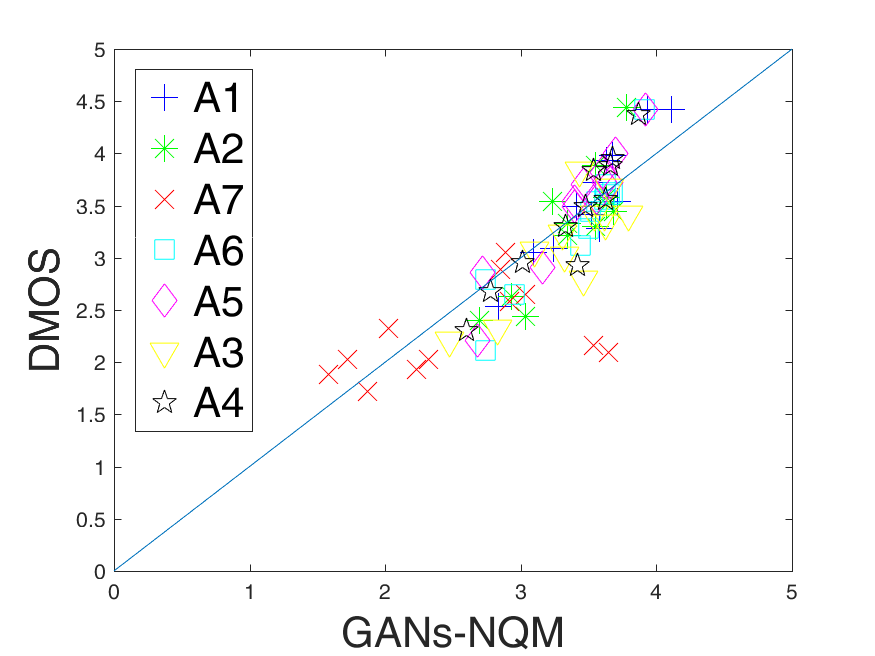}}
 \\ 
 \caption{Scatter plots of objective quality scores versus DMOS on IRCCyN/IVC DIBR database~\cite{bosc2011towards}. A1-A7 represent different DIBR algorithms in \cite{bosc2011towards}. }
 \label{fig:scatter_plot_all_metrics}
\end{figure*}

The scatter plots of all the tested quality metrics versus DMOS are provided in Fig. \ref{fig:scatter_plot_all_metrics}. By comparing the scatter plots of GANs-NQM with others, we can notice that most of the objective scores predicted by the proposed metric are better distributed along the diagonal of the plot. In the scatter plot of APT and NIQSV+, images that synthesized using the same DIBR algorithm are predicted with similar objective scores, which leads to a `vertical line' as shown in Fig. \ref{fig:scatter_plot_apt} and Fig. \ref{fig:scatter_plot_niqsv+} .

\begin{figure*}[!htbp]
\centering
 \mbox{ \parbox{1\textwidth}{
   \begin{minipage}[b]{0.15\textwidth}
   \subfloat[reference ]
  {\label{fig:inpaint_a}\includegraphics[width=\textwidth]{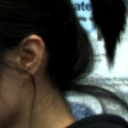}}  
  \end{minipage}
    \begin{minipage}[b]{0.15\textwidth}
  \subfloat[ patch with holes]
  {\label{fig:inpaint_b}\includegraphics[width=\textwidth]{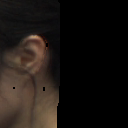}}
  \end{minipage}
   \begin{minipage}[b]{0.15\textwidth}
   \subfloat[ \cite{mueller2009view} PSNR$= 10.5 $ ]
  {\label{fig:inpaint_c}\includegraphics[width=\textwidth]{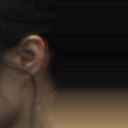}}  
  \end{minipage}
    \begin{minipage}[b]{0.15\textwidth}
  \subfloat[\cite{ndjiki2011depth} PSNR$= 10.4 $  ]
  {\label{fig:inpaint_d}\includegraphics[width=\textwidth]{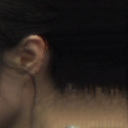}}
  \end{minipage}
    \begin{minipage}[b]{0.15\textwidth}
    \subfloat[\cite{koppel2010temporally} PSNR$= 11.1 $  ]
  {\label{fig:inpaint_e}\includegraphics[width=\textwidth]{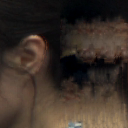}}
   \end{minipage} 
       \begin{minipage}[b]{0.15\textwidth}
    \subfloat[Ours PSNR$= 12.5$   ]
  {\label{fig:inpaint_f}\includegraphics[width=\textwidth]{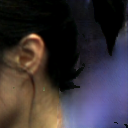}}
   \end{minipage} 
   \\
 \begin{minipage}[b]{0.15\textwidth}
   \subfloat[reference   ]
  {\label{fig:inpaint_g}\includegraphics[width=\textwidth]{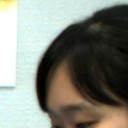}}  
  \end{minipage}
    \begin{minipage}[b]{0.15\textwidth}
  \subfloat[patch with holes ]
  {\label{fig:inpaint_h}\includegraphics[width=\textwidth]{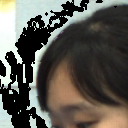}}
  \end{minipage}
    \begin{minipage}[b]{0.15\textwidth}
   \subfloat[\cite{mueller2009view} PSNR$= 24.3 $]
  {\label{fig:inpaint_i}\includegraphics[width=\textwidth]{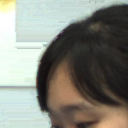}}  
  \end{minipage}
    \begin{minipage}[b]{0.15\textwidth}
  \subfloat[\cite{ndjiki2011depth} PSNR$= 23.9 $ ]
  {\label{fig:inpaint_j}\includegraphics[width=\textwidth]{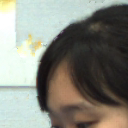}}
  \end{minipage}
    \begin{minipage}[b]{0.15\textwidth}
    \subfloat[\cite{koppel2010temporally} PSNR$= 24.3 $]
  {\label{fig:inpaint_k}\includegraphics[width=\textwidth]{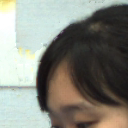}}
   \end{minipage} 
       \begin{minipage}[b]{0.15\textwidth}
    \subfloat[Ours PSNR$= 27.5 $]
  {\label{fig:inpaint_l}\includegraphics[width=\textwidth]{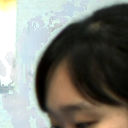}}
   \end{minipage} 
  \\
   \begin{minipage}[b]{0.15\textwidth}
   \subfloat[ reference ]
  {\label{fig:inpaint_m}\includegraphics[width=\textwidth]{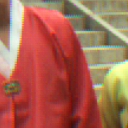}}  
  \end{minipage}
    \begin{minipage}[b]{0.15\textwidth}
  \subfloat[ patch with holes]
  {\label{fig:inpaint_n}\includegraphics[width=\textwidth]{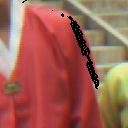}}
  \end{minipage}
   \begin{minipage}[b]{0.15\textwidth}
   \subfloat[\cite{mueller2009view}  PSNR$= 25.4 $ ]
  {\label{fig:inpaint_o}\includegraphics[width=\textwidth]{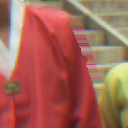}}  
  \end{minipage}
    \begin{minipage}[b]{0.15\textwidth}
  \subfloat[\cite{ndjiki2011depth}  PSNR$= 25.5 $ ]
  {\label{fig:inpaint_p}\includegraphics[width=\textwidth]{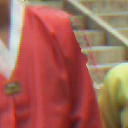}}
  \end{minipage}
    \begin{minipage}[b]{0.15\textwidth}
    \subfloat[\cite{koppel2010temporally} PSNR$= 25.6 $]
  {\label{fig:inpaint_q}\includegraphics[width=\textwidth]{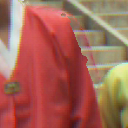}}
   \end{minipage}  
     \begin{minipage}[b]{0.15\textwidth}
    \subfloat[Ours PSNR$=28.1 $ ]
  {\label{fig:inpaint_r}\includegraphics[width=\textwidth]{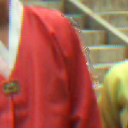}}
   \end{minipage} 
  \\
   \begin{minipage}[b]{0.15\textwidth}
   \subfloat[reference  ]
  {\label{fig:inpaint_s}\includegraphics[width=\textwidth]{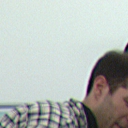}}  
  \end{minipage}
    \begin{minipage}[b]{0.15\textwidth}
  \subfloat[ patch with holes]
  {\label{fig:inpaint_t}\includegraphics[width=\textwidth]{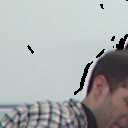}}
  \end{minipage}
      \begin{minipage}[b]{0.15\textwidth}
   \subfloat[\cite{mueller2009view} PSNR$= 23.2$  ]
  {\label{fig:inpaint_u}\includegraphics[width=\textwidth]{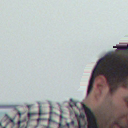}}  
  \end{minipage}
    \begin{minipage}[b]{0.15\textwidth}
  \subfloat[\cite{ndjiki2011depth}  PSNR$= 23.0 $ ]
  {\label{fig:inpaint_v}\includegraphics[width=\textwidth]{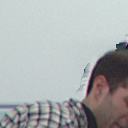}}
  \end{minipage}
    \begin{minipage}[b]{0.15\textwidth}
    \subfloat[\cite{koppel2010temporally} PSNR$= 23.9$  ]
  {\label{fig:inpaint_w}\includegraphics[width=\textwidth]{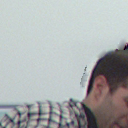}}
   \end{minipage}  
     \begin{minipage}[b]{0.15\textwidth}
   \subfloat[Ours PSNR$= 27.6 $  ]
  {\label{fig:inpaint_x}\includegraphics[width=\textwidth]{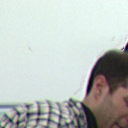}}
   \end{minipage}   
  }}
\caption{Results of using our re-trained generator to inpaint the dis-occluded regions. First column: reference patches; Second column: patches with dis-occluded regions; Third column: inpainted results using algorithm proposed in~\cite{mueller2009view}; Forth column:  inpainted results using algorithm proposed in~\cite{ndjiki2011depth}; Fifth column:inpainted results using algorithm proposed in~\cite{koppel2010temporally}; Sixth column: inpaintd results using our retrained generator;}
\label{fig:example_inpainted}
\end{figure*}

In order to examine the significance of the performances between each two tested quality metrics, Student's t-test is conducted. More specifically, the 1000 PCC values obtained during the cross performance evaluation described in section for each tested metric are used as input for t-test. The results are concluded in Table \ref{tab:signifivance} with a significance level of 0.05, where `1' represents the performance of the under-test metric in row outperforms the one in column significantly, `-1' represents the inverse situation and `0' represents there is no significant difference. According to the table, the performance of the proposed GANs-NQM is significantly better than all the other NR and FR metrics except for ST-IQM, with which it shows no statistically different performance. Considering the fact that our method is NR, the proposed GANs-NQM is more applicable in real scenarios.  
\begin{table*}[!htbp]
\centering
\caption{Statistic significance results based on the 1000 times cross performance evaluation}
\label{tab:signifivance}
\begin{tabular}{|c|c|c|c|c|c|c|c|c|c|c|c|c|}\hline
\multirow{2}{*}{Metric} & 3DS & VS   & MP-PS  & MP-PS & MW-PS & MW-PS & CT- & EM- &  ST- &  NIQ  &  \multirow{2}{*}{APT} & GANs\\ 

 &  wIM  & QA   & NR$_{red}$ & NR$_{ful}$ & NR$_{red}$ & NR$_{ful}$ & IQM & IQM &  IQM & SV+ &   &  -NQM \\ \hline
3DSwIM          & -    & 0               & 0               & 0               & 0               & 0      & 0      & 0      & -1     & 0   & 0        & -1 \\\hline
VSQA            & 0    & -               & 0               & 0               & 0               & 0      & 0      & -1     & -1     & 0   & -1       & -1 \\\hline


MP-PSNR$_{red}$ & 0    & 0               & -               & 0               & 0               & 0      & 0      & 0      & 0      & 0   & 0        & -1 \\\hline
MP-PSNR$_{ful}$ & 0    & 0               & 0               & -               & 0               & 0      & 0      & 0      & 0      & 0   & 0        & -1 \\\hline
MW-PSNR$_{red}$ & 0    & 0               & 0               & 0               & -               & 0      & 0      & 0      & -1     & 0   & 0        & -1 \\\hline
MW-PSNR$_{ful}$ & 0    & 0               & 0               & 0               & 0               & -      & 0      & 0      & -1     & 0   & 0        & -1 \\\hline
CT-IQM          & 0    & 0               & 0               & 0               & 0               & 0      & -      & 0      & 0      & 0   & 0        & -1 \\\hline
EM-IQM          & 0    & 0               & 0               & 0               & 0               & 0      & 0      & -      & 0      & 0   & 0        & -1 \\\hline
ST-IQM          & 1    & 1               & 0               & 0               & 1               & 1      & 0      & 0      & -      & 1   & 0        & 0  \\\hline
NIQSV+          & 0    & 0               & 0               & 0               & 0               & 0      & 0      & 0      & -1     & -   & 0        & -1 \\\hline
APT             & 0    & 0               & 0               & 0               & 0               & 0      & 0      & 0      & 0      & 0   & -        & -1 \\\hline
GANs-NQM        & 1    & 1               & 1               & 1               & 1               & 1      & 1      & 1      & 0      & 1   & 1        & - \\ \hline

\end{tabular}
\end{table*}

\begin{table*}[!htbp]
\centering
\caption{ Normalized execution time of each metric  }
\label{tab:runtime}
\begin{tabular}{|c|c|c|c|c|c|c|c|c|c|c|c|c|}\hline
 & \multicolumn{9}{c|}{FR Metrics } &  \multicolumn{3}{c|}{NR Metrics } \\ \hline

\multirow{2}{*}{Metric} & MW-PS & MW-PS & MP-PS  & 3DS & MW-PS &  \multirow{2}{*}{Logs} & EM- & CT- &  ST- &  NIQ  & GANs &  \multirow{2}{*}{APT}\\ 

 &   NR$_{red}$  & NR$_{ful}$  & NR$_{red}$ &  wIM &  NR$_{ful}$  &    & IQM & IQM &  IQM & SV+ &  -NQM  &  \\ \hline

 Time & 9.6 & 12.4 & 35 & 90 & 100& 220.4 &  1.3k+ & 4.5k+  & 12.7k+ & 21   &  \textbf{157}  & 13k+ \\ \hline

\end{tabular}
\end{table*}

Another important application of an objective metric in FTV system is to benchmark different synthesized algorithms, the ground truth ranking of the seven synthesis algorithms (A1-A7) used in our study is obtained by averaging the DMOS. The predicted rankings by objective metrics are reported in Table \ref{tab:Rank}. According to Table \ref{tab:Rank}, the ranking of the proposed GANs-NQM is the most consistent. For GANs-NQM, only the rankings of A4 and A5, which generate similar quality synthesized images, are switched. Therefore, a desirable rank order could be provided by the proposed model to select proper synthesis algorithms. 


\begin{table}[!htbp]
\begin{center}
\caption{\label{tab:Rank}%
Ranking of RGB-D synthesis algorithms} 
{
\renewcommand{\baselinestretch}{1}\footnotesize
\begin{tabular}{|c|c|c|c|c|c|c|c|}
\hline
 & \multicolumn{7}{ c| }{ Ranking of synthesis algorithm}\\ \hline
  DMOS (ground truth) &\bf{A1} &\bf{A5} &\bf{A4} &\bf{A6} &\bf{A2} &\bf{A3} &\bf{A7}        \\ \hline
 \multicolumn{8}{ |c| }{ Full Reference Metric (FR)}\\ \hline
 3DSwIM & A1 & A4 & A5& A6 & A3& A2 & A7\\ \hline
 VSQA & A6 & A5 & A4 & A3& A2 & A7  & A1\\ \hline
 MP-PSNR$_{red}$ & A4 & A5 & A6 & A3 & A2 & A1 & A7\\ \hline
 MP-PSNR$_{ful}$ & A4 & A5 & A6 & A3 & A2 & A1 & A7\\ \hline
 MW-PSNR$_{red}$  &A4 & A5 & A6 & A2 & A3 & A1 & A7\\ \hline
 MW-PSNR$_{ful}$  & A4 & A5 & A6 & A2 & A3 & A1 & A7\\ \hline
 CT-IQM &  A1  & A4  & A2 &  A5  & A6 & A3  & A7\\ \hline
 EM-IQM  &  A1  & A2  & A6  &  A3  & A4 & A5  &  A7\\ \hline
 ST-IQM  &  A1 &  A5  & A6 &  A4 &  A3 & A2  & A7 \\ \hline
 \multicolumn{8}{ |c| }{NO Reference Metric (NR)}\\ \hline

NIQSV+ & A1 & A6 & A5 & A4 & A2 & A3 & A7\\ \hline
APT  & A1 & A2 & A4 & A3 & A5 & A6 & A7\\ \hline
GANs-NQM &  A1 & A4 & A5 & A6 & A2 & A3 & A7\\ \hline
\end{tabular}}
\end{center}
\end{table}

Last but not least, to make the evaluation of the quality of RGB-D synthesized view feasible in real applications, the time cost of the quality assessment metric should be reasonable, if possible, the lower the better.  To verify the efficiency of the proposed metric, as well as make comparisons with others, the execution time normalized by the run time of PSNR is computed~\cite{tian2018niqsv+}. By calculating the normalized execution time, it is then possible to compare the time complexities of different metrics on different machines and datasets. For a given image $x$ from a database, the normalized execution time $t_{norm}$ is defined as  
\begin{equation} 
t_{norm}= \frac{t_{metric}}{t_{PSNR}} ,
\end{equation} 
 where $t_{metric}$ is the execution time of the objective quality metric for image $x$, and $t_{PSNR}$ is the corresponding runtime of PSNR. For completeness, in our study, the experiments are conducted on a desktop equipped with i7 CPU (4GHz), 8 GB RAM, and a Nvidia Xeon E3-1200 v3/4th. The runtime of PSNR for one synthesized image in IRCCyN/IVC DIBR images database is 0.05 seconds. The normalized execution time for each metric is summerized in Table~\ref{tab:runtime}. Although GANs-NQM is slower than NIQSV+, it is still within a reasonable time cost, which is much faster than the second best NR metric APT, as well as the best FR metric ST-IQM.

\subsection{Inpainting results}
The theoretical assumption of this paper is that the generator/discriminator are simultaneously trained to inpaint/evaluate the RGB-D dis-occluded regions/RGB-D synthesis views. The performance of utilizing discriminator to predict the quality of the RGB-D synthesis views has been demonstrated in the previous section. As a side outcome of this paper, it would be interesting to evaluate the performance of the pre-trained context inpainter (generator) on the same database, \textit{i.e.}, the synthesized views that contain dis-occluded regions in the IRCCyN/IVC DIBR images database.

PSNR between the reference and the inpainted image is calculated for evaluation. Three inpainting algorithms~\cite{ndjiki2011depth}~\cite{mueller2009view}~\cite{koppel2010temporally} are used for comparison. Due to the limitation of space, selected results are shown in Fig.\ref{fig:example_inpainted}. 

Based on the results, it is observed that 
1) By comparing our inpainted result in Fig.~\ref{fig:inpaint_f} to the others with respect to the reference, the shape of the braid of the girl is better remained by our model. Similar results could also be observed in Fig.~\ref{fig:inpaint_l} where the corner of the poster is better preserved compared to the others; 
2) The shape of the dis-occluded regions in Fig.~\ref{fig:inpaint_n} are better inpainted by the proposed models as shown in Fig.~\ref{fig:inpaint_r}. There are obvious `double-edge like' shapes, \textit{i.e.} ghosting artifacts, along the objects after being inpainted by other methods; 
3) In the condition that holes appear in homogeneous texture regions which are also close to the borders of foreground objects, our inpainted result is with higher texture consistency than the others as shown in Fig.~\ref{fig:inpaint_x}.

In conclusion, our proposed context inpainter could maintain the structures of the dis-occluded regions, especially when the dis-occluded regions are large. For the challenging dis-occluded regions that lie on the border of foregrounds and backgrounds, as well as in the homogeneous texture regions close to the border of foreground objects, the proposed inpainter performs better than the others. 

The appealing performance of our pre-trained context inpainter (generator) on RGB-D dis-occluded regions validates the effectiveness of the proposed training strategy, which uses specific designed masks to mimic the typical black-hole artifacts induced in DIBR process. The proposed strategy is more flexible in using the large-scale image databases in the computer vision domain rather than the RGB-D datasets where the depth information might be noisy. It should also be noted that the training data scale in our study is only 10K, which could be definitely further augmented by employing the existing datasets. Therefore, there is still improvement space for our current trained model, no matter for quality assessment or for hole filling of RGB-D synthesis view.



\section{Conclusions}
\label{sec:con}

In this work, we proposed a GANs-based NR quality metric, GANs-NQM, to evaluate the perceptual quality of RGB-D synthesis views. To resolve the challenges of the training data scales in DNNs, a novel strategy is proposed which exploits the current existing large-scale 2D computer vision datasets rather than RGB-D datasets, where depth data may be unreliable. The spirit of the strategy can be easily applied to other applications in RGB-D domain or even other community. Based on the assumption that if a generator of a GANs could be trained to inpaint the dis-occluded regions then the discriminator could be used to predict the quality, in this study, we learned a `Bag of Distortions Word' (BDW) codebook, proposed a local distortion region selector from the discriminator, and eventually mapped the non-uniform inpainting related artifacts to perceptual quality through SVR. According to experimental results, the proposed GANs-NQM provides the best performance compared to the state-of-the-art FR/NR quality metrics for RGB-D synthesized views. As a side outcome, the pre-trained inpainter also shows an appealing performance in inpainting the challenging holes in RGB-D synthesis view.

\bibliographystyle{IEEEtran}
\bibliography{bare_jrnl_compsoc.bbl}

\end{document}